\documentclass{emulateapj}
\slugcomment{To Appear in ApJ}

\shorttitle{SNRs AND TYPE Ia SN PROGENITORS}
\shortauthors{Badenes et al.}

\begin{document}

\title{ARE THE MODELS FOR TYPE Ia SUPERNOVA PROGENITORS CONSISTENT WITH THE PROPERTIES OF SUPERNOVA REMNANTS?}

\author{Carles Badenes\altaffilmark{1,2}, John P. Hughes\altaffilmark{1,3}, Eduardo Bravo\altaffilmark{4}, and Norbert
  Langer\altaffilmark{5}}

\altaffiltext{1}{Department of Physics and Astronomy, Rutgers University, 136 Frelinghuysen Rd., Piscataway NJ
  08854-8019; badenes@physics.rutgers.edu, jph@physics.rutgers.edu}

\altaffiltext{2}{\textit{Chandra} Fellow}

\altaffiltext{3}{Department of Astrophysical Sciences, Princeton University, Princeton, NJ 08544}

\altaffiltext{4}{Departament de F\'{i}sica i Enginyeria Nuclear, Universitat Polit\`{e}cnica de Catalunya, Diagonal 647,
  Barcelona 08028, Spain; and Institut d'Estudis Espacials de Catalunya, Campus UAB, Facultat de Ci\`{e}ncies. Torre
  C5. Bellaterra, Barcelona 08193, Spain; eduardo.bravo@upc.es}

\altaffiltext{5}{Astronomical Institute, University Utrecht, P.O. Box 80000, 3508TA Utrecht, the Netherlands;
  N.Langer@astro.uu.nl}

\begin{abstract}
  We explore the relationship between the models for progenitor systems of Type Ia supernovae and the properties of the
  supernova remnants that evolve after the explosion. Most models for Type Ia progenitors in the single degenerate
  scenario predict substantial outflows during the presupernova evolution. Expanding on previous work, we estimate the
  imprint of these outflows on the structure of the circumstellar medium at the time of the supernova explosion, and the
  effect that this modified circumstellar medium has on the evolution of the ensuing supernova remnant. We compare our
  simulations with the observational properties of known Type Ia supernova remnants in the Galaxy (Kepler, Tycho, SN
  1006), the Large Magellanic Cloud (0509-67.5, 0519-69.0, N103B), and M31 (SN 1885). We find that optically thick
  outflows from the white dwarf surface (sometimes known as `accretion winds') with velocities above $200 \, \mathrm{km
    \, s^{-1}}$ excavate large low-density cavities around the progenitors. Such large cavities are incompatible with
  the dynamics of the forward shock and the X-ray emission from the shocked ejecta in all the Type Ia remnants that we
  have examined.
\end{abstract}


\section{INTRODUCTION}
\label{sec:Intro}

Despite decades of continuing effort, the progenitor systems of Type Ia supernovae (SNe) have not been confidently
identified yet \citep[see][and references therein]{hillebrandt00:Ia-review}. This has become one of the most pressing
problems in stellar evolution, with profound implications for cosmology and the chemical evolution of galaxies
\citep{kobayashi98:lowZ_inhibition_IaSNe}. It has been known for some time that Type Ia SNe are the result of the
thermonuclear explosion of a C+O white dwarf (WD) that is destabilized by accretion of matter from a companion in a
close binary system \citep{hoyle60:nucleosynth-SNe}, but the details of this process are still the subject of heated
debate. In particular, the nature of the companion and the physical processes involved in the evolution of the binary
prior to explosion remain obscure.

Most of the discussions on Type Ia progenitor systems are held in the framework of two basic scenarios. In the so-called
single degenerate (SD) scenario, the WD companion is a normal or slightly evolved star. In the double degenerate (DD)
scenario, it is another WD. In the SD scenario, the WD might explode either close to the Chandrasekhar limit (SD-Ch) or
some time before reaching it (sub-Chandrasekhar explosions, SD-subCh). Several observational and theoretical arguments
point at SD-Ch explosions as the most promising route to Type Ia SNe, but there are still many uncertainties associated
with this paradigm \citep[for discussions, see][]{branch95:Ia-prog,tout05:Type_Ia_Progenitors}. The main problem is the
necessity to bring the mass of the C+O WD close enough to the Chandrasekhar limit to trigger the explosion. At the onset
of the final mass transfer episode in SD systems, the WD mass is thought to lie between 0.7 $\mathrm{M_{\odot}}$
\citep{langer00:prog-models,han04:SDchannel_for_SNIa} and 1.2 $\mathrm{M_{\odot}}$ \citep{nomoto82:accretingWD}. This
means that, in order to achieve the $\sim 1.38 \, \mathrm{M_{\odot}}$ necessary for a thermonuclear runaway, the WD
needs to accrete between 0.2 and 0.7 $\mathrm{M_{\odot}}$ of material from its companion and then burn it somehow to C
and O avoiding the highly unstable degenerate conditions at the WD surface. Our present understanding of Type Ia SNe in
the SD-Ch scenario is based on theoretical models that attempt to explain the details of these accretion and burning
processes.

One key aspect of such models is the presence or absence of outflows from the binary system during its pre-SN
evolution. If these outflows are indeed present, they should leave some kind of imprint on the circumstellar medium
(CSM) of the progenitor, but the evidence for this modified CSM in Type Ia SN observations remains inconclusive. Prompt
emission at radio or X-ray wavelengths from the interaction of SN ejecta with surrounding material, which has been found
in several core collapse SNe, appears to be absent in Type Ia events. In the radio, \citet{panagia06:radio_SNIa} report
no VLA detections in 27 nearby objects. In the X-rays, no Type Ia SNe have been detected with \textit{ROSAT}
\citep[upper limits on one object,][]{schlegel93:xraycsm} or \textit{Chandra} (upper limits on three objects, Hughes et
al. in preparation). \citet{immler06:Swift_Ia_SN2005ke} report \textit{Swift} observations of eight Type Ia SNe, with
only one marginal ($\sim 3 \sigma$) detection, SN 2005ke, a peculiar object that was not seen by \textit{Chandra}. It is
important to keep in mind that these studies only probe the CSM at distances between $10^{15}\,\mathrm{cm}$ (for
observations weeks or months after the explosion) and $10^{17}\,\mathrm{cm}$ \citep[for the latest radio observations
in] [several years after the explosion]{panagia06:radio_SNIa}. Modification of the CSM on larger scales will become
apparent in the aftermath of the explosion, during the supernova remnant (SNR) phase. In \citet{badenes01:imprint}
(henceforth BB01), we conducted a preliminary search for the imprint of presupernova outflows on the dynamics of young
Type Ia SNRs. In the present work, we expand and revise the calculations presented in BB01, and we perform a more
detailed and robust comparison to the observations of SNRs. Our aim is to use these comparisons as a means to constrain
the fundamental properties of the presupernova outflows and gain insights on the nature of the progenitor systems.

This paper is organized as follows. First, we examine the outflow mechanisms from Type Ia progenitors in the SD-Ch
channel, and we build simple models to mimic them (\S~\ref{sec:outflows}). From these outflow models, we simulate the
structure of the CSM around the progenitor system at the time of the SN explosion (\S~\ref{sec:CSM}) and the subsequent
evolution of the SNR (\S~\ref{sec:dynamics-snr}). In \S~\ref{sec:comp-with-obs-SNRs}, we compare our SNR models with the
known properties of the forward shock dynamics and X-ray emission in a sample of known Type Ia SNRs. Finally, we discuss
our results in \S~\ref{sec:Discussion} and we summarize them in \S~\ref{sec:Conclusions}.

\section{TYPE Ia PROGENITORS IN THE SINGLE DEGENERATE CHANNEL: PRESUPERNOVA OUTFLOWS}
\label{sec:outflows}

\subsection{Pre-SN Evolution Scenarios}
\label{sub:outflows-overview}

There are two channels by which a SD binary system can lead to a Type Ia SN. The systems in the first channel consist of
a WD and a main sequence (MS) or slightly evolved companion star \citep[WD+MS
channel,][]{li97:supersoft,hachisu99:supersoft}. Mass transfer occurs via Roche lobe overflow, which requires that the
two components be very close, with an initial period of the order of days. A different class of Type Ia progenitors
systems is composed of a WD and a red giant (RG) star with a wider separation and a longer initial period, of the order
of a hundred days \citep[WD+RG channel,][]{li97:supersoft,hachisu99:symbiotic}. The mechanism for mass transfer in these
more detached systems is not well understood, but Roche lobe overflow might still play a significant role.

The details of the evolution of Type Ia progenitor systems in these two channels are very complex. This is mainly due to
the fact that the process of mass transfer affects both the orbital parameters of the system and the evolution of the
components, which in turn modify the conditions for the mass transfer. In the most widely accepted picture for SD Type
Ia progenitors, the H- or He-rich material from the donor star forms an accretion disk and then accumulates in a shell
on top of the WD. Given the appropriate conditions, nuclear burning in this shell converts the accreted material to C
and O, until the mass of the WD reaches $\sim1.38\,\mathrm{M_{\odot}}$. At this point, a runaway thermonuclear flame
ignites in the central regions of the WD, and a Type Ia SN explosion ensues. It is important to stress that a SN
explosion is only one of the many possible outcomes of the evolution of close binary systems that contain a WD. The key
parameter that determines whether a specific binary system leads to a Type Ia SN or not is the accretion rate of
the WD, $\dot{M}$.

For accretion rates below a critical limit $\dot{M}_{stable}$, nuclear burning on the surface of the degenerate WD is
unstable \citep{nomoto82:accretingWD} \footnote{In a recent paper, \citet{starrfield04:SS_Ia} proposed a stable
  accretion regime for low values of $\dot{M}$, but see \citet{nomoto06:Thermal_Stability_H_Accreting_WDs}}. The value
of $\dot{M}_{stable}$ depends on the mass of the WD \citep{nomoto91:AIC}, on the metallicity of the accreted material
\citep{piersanti00:H_accreting_CO_WD_II}, and on the rotation of the WD
\citep{uenishi03:rotating_WDs_Ia,yoon04:WD_rotation_He_shell_burning,yoon04:thermonuclear_shell_stability}, but it is
generally assumed to be of the order of $10^{-7}\,\mathrm{M_{\odot} \, yr^{-1}}$. If $\dot{M}$ is only slightly lower
than $\dot{M}_{stable}$, nuclear burning proceeds cyclically in weak flashes, and most of the accumulated material might
be retained by the WD \citep{kato04:mass_accumulation_He_flashes}. Recurrent novae like RS Oph or U Sco are thought to
be the observational counterparts of white dwarf binaries accreting in this regime
\citep{hachisu01:recurrent_novae}. For even lower values of $\dot{M}$, however, the flashes become stronger, effectively
turning into nova explosions that erode the WD \citep{yaron05:nova_models_2}. If the accretion enters this regime, the
WD mass effectively decreases with time and the system cannot produce a Type Ia SN \citep[see Fig. 5 in][and references
therein]{kahabka97:SSXS_review}.

At higher accretion rates, the energy input from the shell nuclear burning can lead to the expansion of the outer layers
of the WD to red giant dimensions \citep{nomoto91:AIC}. This occurs when $\dot{M}$ approaches the nuclear
Eddington-accretion rate $\dot M_{Edd,nuc} = L_{Edd} / \varepsilon_{nuc}$, where $L_{Edd}= 4\pi c G M / \kappa$ is the
Eddington luminosity of the WD, and $\varepsilon_{nuc} \simeq 7 \times 10^{18}\,\mathrm{erg\, g^{-1}}$ is the energy
gained by fusing H into He. For a WD of one solar mass, $\dot M_{\rm Edd,nuc} \simeq 6 \times
10^{-7}\,\mathrm{M_{\odot}\,yr^{-1}}$.  Upon expansion of the WD, the binary system may enter a common envelope phase,
which might lead to a spiral-in (due to dynamical friction) and the merging of both components
\citep{iben93:common_envelope,ivanova04:mass_transfer_WD_binaries}. For a long time, it was thought that this would
prevent many SD systems from becoming Type Ia progenitors, but \citet{hachisu96:progenitors} (henceforth, HKN96)
proposed a mechanism to avoid the common envelope phase. These authors claimed that, for high values of $\dot{M}$, the
luminosity from the shell nuclear burning can drive an optically thick `accretion wind' outflow from the WD surface with
an outflow mass loss rate $\dot{M}_{of}$ such that the effective value of $\dot{M}$ remains below
$\dot{M}_{Edd,nuc}$. This accretion wind results in the loss of material from the binary system.

\subsection{Accretion Wind Outflows from Type Ia Progenitor Systems}
\label{sub:outflows-outflows}

In recent times, most of the models for SD Type Ia SN progenitors have relied to a greater or lesser extent on the
accretion wind mechanism of HKN96 as a means to limit the effective value of $\dot{M}$. This is despite the fact that
several fundamental aspects of the accretion wind mechanism are not well understood. In particular, some kind of
self-regulation is required so that $\dot{M}$ falls below $\dot{M}_{Edd,nuc}$ (to avoid the expansion of the WD), but at
the same time remains very close to this critical value (in order for the WD to gain enough mass to explode as a Type Ia
SN). Furthermore, the accretion wind is supposed to remain active up to mass transfer rates of the order of the
Eddington accretion rate $\dot M_{Edd,acc} = L_{Edd} / \varepsilon_{acc}$, where $\varepsilon_{acc} = GM/R$ is the
specific potential energy gained by accreting matter onto a WD with radius $R$. In this situation, $\dot{M}_{of} \simeq
\dot M_{Edd,acc}$, which implies, for wind velocities similar to the WD escape velocity (see
\S~\ref{sub:outflows-theory}), a ratio of wind-to-photon momentum of $\eta = 1/c \sqrt{2R\kappa/(GM)} \sim 10^{2}$. It
is known that winds driven by photon scattering in lines cannot work in this regime \citep{lucy93:WR_winds}. In
principle, continuum-driven winds might reach such high efficiencies, as claimed by HKN96 \citep[see
also][]{owocki04:supereddington_mass_loss,smith06:continuum_driven_mass_loss}, but this has never been proved with
detailed simulations. Due to these and other concerns, the role that accretion winds play in the presupernova evolution
of Type Ia progenitors remains controversial. Some authors claim that this mechanism cannot work in a realistic
scenario, and that H-accreting WDs cannot explode as Type Ia SNe
\citep{cassisi98:H-accreting-CO-WD,piersanti99:H_accreting_CO_WD_I,piersanti00:H_accreting_CO_WD_II}.

\subsubsection{Theoretical studies}
\label{sub:outflows-theory}

Following the work of HKN96, \citet{hachisu99:supersoft} explored several specific cases within the WD+MS channel,
making use of analytical formulae to follow the evolution of the binary system parameters and mass transfer rates. In a
companion paper, \citet{hachisu99:symbiotic} applied the same techniques to systems in the WD+RG channel. In these
papers, substantial outflows from the Type Ia progenitors were found in all cases. The typical timescales for the
outflows are several times $10^{5}$ yr, with mass loss rates $\dot{M}_{of}$ of the order of $10^{-6}\,\mathrm{M_{\odot}
  \, yr^{-1}}$. In some binary systems, the WD explodes as a Type Ia SN while the optically thick accretion wind is
still active. In other cases, the outflow stops some time before the explosion, leading to a more or less extended
mass-conservative phase.

Other studies have followed the evolution of binary systems with accretion winds, relying on stellar evolution codes
instead of the analytical formulae used by \citet{hachisu99:supersoft,hachisu99:symbiotic}. This kind of calculation is
necessary in order to estimate the mass-transfer rates in the system correctly. In general, stellar evolution codes
predict lower values of $\dot{M}_{of}$ and hence longer evolutionary timescales than those obtained using analytical
formulae. The first such study was published by \citet{li97:supersoft}, but a more detailed and updated exploration of
the parameter space (with $\sim 2300$ binary systems) can be found in \citet{han04:SDchannel_for_SNIa}. These works have
in common that they found substantial outflows from the Type Ia progenitors in all cases. In contrast,
\citet{langer00:prog-models} only found outflows in about one third of the progenitor systems of solar metallicity that
they simulated. This is due to more optimistic assumptions concerning the mass accumulation efficiency in accreting WDs,
which might be justified by models that include the effect of rotation \citep{yoon04:WD_rotation_He_shell_burning}. It
is also important to mention that \citeauthor{langer00:prog-models} stopped their simulations for systems which required
outflows with $\dot M_{of} > 3 \dot M_{Edd, nuc}$.

In the theoretical framework laid down by HKN96, accretion winds are formed in the WD envelope with photospheric
velocities of $\sim1000\,\mathrm{km \, s^{-1}}$ \citep{hachisu99:supersoft,hachisu99:symbiotic}. These high outflow
velocities are necessary because the escape velocity from the surface of the WD is also very high \citep[roughly of the
same order of magnitude, see Figure 4 and accompanying discussion
in][]{kato94:optically_thick_winds_nova_outbursts}.  The orbital velocities in Type Ia progenitor systems are much lower
than this, of the order of $200\,\mathrm{km \,s^{-1}}$ \citep[][]{langer00:prog-models}.

\subsubsection{Observations}
\label{sub:outflows-observations}

It is useful to compare the theoretical accretion wind models described in the previous section to the observations of
binary systems that have been proposed as Type Ia progenitor candidates. The list of candidate objects is very
heterogeneous, and includes transient supersoft X-ray sources like RX J0513.9-6951 \citep{hachisu03:rxj0513.9-6951},
eclipsing binaries with intermittent X-ray emission like V Sagittae \citep{hachisu03:VSge}, and recurrent novae like RS
Oph \citep{hachisu01:recurrent_novae} and U Sco \citep{hachisu00:USco}.

Among these objects, RXJ0513.9-6951 and V Sagittae are of particular interest, because they have non-explosive outflows
that can be modeled using the accretion wind mechanism \citep{hachisu03:rxj0513.9-6951,hachisu03:VSge}. In the case of
RXJ0513.9-6951, the outflows have a velocity of $\sim3800\,\mathrm{km \, s^{-1}}$, with strong hints of a bipolar
structure \citep{pakull93:RXJ0153,hutchings02:RXJ0513}. The outflows from V Sagittae are more complex, but a high
velocity component at $\sim1500\,\mathrm{km \, s^{-1}}$ also appears to be present \citep{wood00:VSagittae}. These high
outflow velocities support the theoretical values discussed in the previous Section. It is interesting to note that both
systems are variable sources at optical and X-ray wavelengths, with periods of the order of a few hundred
days. \citet{hachisu03:rxj0513.9-6951} attribute the variability of RXJ0513.9-6951 to the impact of the strong
($\sim2\times10^{-6}\,\mathrm{M_{\odot} \, yr^{-1}}$) accretion winds from the WD atmosphere on the surface of the donor
star, whose outer envelope is stripped off, temporarily shutting down the accretion process and interrupting the
accretion wind itself. The same authors propose a somewhat similar model for V Sagittae, but it is unclear whether this
kind of variability should be ubiquitous among binary systems with accretion winds.

\subsection{Modeling the Outflows}
\label{sub:outflows-models-outflows}

Based on the theoretical studies described in \S~\ref{sub:outflows-theory}, we have produced several models for the
outflows from Type Ia SN progenitor systems. The temporal evolution of the outflow mass loss rate $\dot{M}_{of}$ is
shown in Figure \ref{fig-1}, and the fundamental properties of the models are listed in Table \ref{tab-1}. Model H1
mimics the mass loss of the example system discussed in \S~3.5 of \cite{hachisu99:supersoft} (see their Fig. 7). Model
LV1 is built after the example in Figure 1 of \citet{li97:supersoft}. Models HP1, HP2, and HP3 are approximations to the
outflows from the three representative Type Ia progenitor systems discussed in \S~3 of \cite{han04:SDchannel_for_SNIa}
(Fig. 1a for model HP1, Fig. 1c for model HP2, and Fig. 1e for model HP3). Finally, models L1 and L2 correspond to the
binary systems numbers 2 and 31 in Table 2 of \citet{langer00:prog-models} \citep[see Fig. 7 for model L1 and Fig. 34
in][for model L2]{deutschmann98:Iaprogenitors}.

These outflow models are chosen as representative examples of the typical evolution of Type Ia progenitors of solar
metallicity. With them, we expand and improve on the work presented in BB01, aiming at a more complete exploration of
the parameter space for Type Ia progenitor systems. Our outflow models are evidently very simplified, but we believe
that they capture the essence of the relationship established between the current models for Type Ia progenitor systems
and the CSM (a more detailed discussion of some of the issues concerning the outflow models is deferred to
\S~\ref{sub:Discuss-initial_models}). We note that, although we have sampled binary system calculations with different
initial conditions and from different authors, the time scales and mass loss rates of all the outflows are very similar,
and should shape the CSM and influence the evolution of the SNRs in a similar way.

\section{THE SHAPING OF THE CIRCUMSTELLAR MEDIUM}
\label{sec:CSM}

\subsection{Method and Parameters}
\label{sub:CSM-method}

We have performed hydrodynamic simulations to calculate the impact that the outflow models presented in
\S~\ref{sub:outflows-models-outflows} would have on the structure of the CSM around the Type Ia progenitors at the time
of the SN explosion. For this purpose, we have used VH-1, a numerical hydrodynamics code developed at the University of
Virginia by John Hawley, John Blondin and collaborators. VH-1 is based on the piecewise parabolic method of
\citet{colella84:PPM}, and the code has been extensively tested and validated in multiple astrophysical scenarios. For
the timescales and velocities involved in the outflows from Type Ia SN progenitors, radiative cooling is dynamically
important. We have included this process through the standard cooling curves of \citet{sutherland93:cooling} for
temperatures above $10^{4}\,\mathrm{K}$, and the values of \citet{dalgarno75:Heating_Ionization_HI} for
$10^{2}<T<10^{4}\,\mathrm{K}$ \citep[for a discussion of the cooling curves,
see][]{raga97:bow_shock_cooling_ionization}. The simulations are performed in one dimension with spherical symmetry, and
they do not include thermal conduction.

Once a particular outflow model is selected, the only parameters that need to be adjusted in our calculations are the
outflow velocity $u_{of}$, and the density and temperature of the interstellar medium (ISM), $\rho_{ISM}$ and
$T_{ISM}$. Following the discussion in \S~\ref{sub:outflows-theory} and \S~\ref{sub:outflows-observations}, we have
chosen a reference value of $1000\,\mathrm{km \, s^{-1}}$ for $u_{of}$. For the ISM parameters $\rho_{ISM}$ and
$T_{ISM}$ we have chosen reference values typical of the warm atomic phase of the ISM:
$\rho_{ISM}=10^{-24}\,\mathrm{g\,cm^{-3}}$ and $T_{ISM}=10^{4}\,\mathrm{K}$ \citep{ferriere01:ISM}. After a brief
theoretical discussion (\S~\ref{sub:CSM-theory}), we will present the CSM structures generated by the outflow models
using these fiducial values (\S~\ref{sub:CSM-results-reference}), and then we will examine the consequences of varying
them (\S~\ref{sub:CSM-results-modified}).

\subsection{Theory}
\label{sub:CSM-theory}

A complete theoretical overview of the interaction between stellar winds and the surrounding medium is given by
\citet{koo92:bubbles_I,koo92:bubbles_II}, who built on the pioneering work of \citet{castor75:interstellar_bubbles} and
\citet{weaver77:interstellar_bubbles_II}. In the framework set by this theory, we can assimilate the accretion wind
outflows to stellar winds, and expect them to form cavities or `accretion wind-blown bubbles' around the Type Ia
progenitors, similar to the more familiar stellar wind-blown bubbles found around the massive progenitors of
core-collapse SNe \citep[see][and references therein]{dwarkadas05:SNR-Bubbles_1D}. In general, these cavities will have
the usual structure of two shocks and a contact discontinuity (CD) separating four regions from the inside out: freely
expanding outflow, shocked outflow, shocked ISM, and unperturbed ISM. The detailed configuration and radial extent of
the cavity will depend on the temporal evolution of the outflow mechanical luminosity
$L_{of}=(1/2)\dot{M}_{of}u_{of}^{2}$, the duration of the mass loss episode, and the pressure exerted by the ISM
\citep[for a review of all the possibilities, see \S~7 in][]{koo92:bubbles_II}. One parameter that is fundamental for
the evolution of the CSM structure is the critical outflow velocity $u_{cr}$, given by

\begin{equation}
u_{cr} = 10^{4} \left\lbrack  \frac{\dot{M}_{of} u_{of}^{2}}{2} \frac{\rho_{ISM}}{\mu_{H}} \right\rbrack ^{1/11} \;\mathrm{cm\,s^{-1}}
\end{equation}

where $\mu_{H}=2.34 \times 10^{-24} \, \mathrm{g}$ is the mean mass per H atom in a solar abundance gas and all
magnitudes are in c.g.s. units \citep[expression adapted from eq. 2.5 in][]{koo92:bubbles_I}. Outflows with $u_{of} >
u_{cr}$ are `fast', meaning that radiative losses do not affect the shocked outflow, and the expansion of the
outer shock is driven by the thermal energy of the shocked material inside the cavity (energy-driven bubbles). Outflows
with $u_{of} < u_{cr}$ are `slow', and radiative losses affect the shocked outflow, either at the reverse shock or at
the CD. These cavities are driven by the ram pressure of the outflow itself (momentum-driven bubbles). For the values of
$\dot{M_{of}}$ and $\rho_{ISM}$ typical of accretion winds expanding into the warm ISM, the outflows are in the fast
regime at velocities above $\sim 100\,\mathrm{km \, s^{-1}}$ (see Figure \ref{fig-2}).

\subsection{Results: Reference Values}
\label{sub:CSM-results-reference}

The structure of the CSM at the time of the SN explosion for each outflow model at the reference values of $u_{of}$,
$\rho_{ISM}$, and $T_{ISM}$ is shown in Figure \ref{fig-3}. The fundamental properties of the CSM profiles are
summarized in Table \ref{tab-2}. As expected for outflows in the fast regime with timescales of $\sim 10^{6}$ yr, all
the accretion winds excavate very large low-density cavities around the Type Ia progenitor systems, with radii $R_{c}$
between 17 and 36 pc. The temporal evolution of these cavities can be divided into two distinct phases. During the first
phase, the outer shock is radiative, and the bubble expands at supersonic velocities. Thin shells of radiatively cooled
material develop behind the outer shock, and then become thicker as the temperature of the material drops to $10^{2}$ K
and cooling ceases. In the terminology introduced by \citet{koo92:bubbles_II}, this is an adiabatic bubble with a
radiative outer shock (ABROS), but the only accretion wind-blown cavity that is still in this stage at $t_{SN}$ is that
generated model HP3. Eventually, the pressure inside the cavity drops below the ISM pressure (which is
$p_{ISM}=\rho_{ISM}T_{ISM}k/\bar{\mu}=1.33\times10^{-12}\,\mathrm{dyn\,cm^{-2}}$ for the reference ISM parameters), and
the cavity becomes pressure-confined. During this stage, the outer shock becomes a standard sound wave that rapidly
dissipates, and the bubble expands (or contracts) at subsonic speeds \citep[adiabatic pressure confined bubble, APCB,
see \S~4.4 in][]{koo92:bubbles_II}. Models H1, LV1, L1, L2, HP1, and HP2 are in this stage at $t_{SN}$. The reverse
shock can only be seen in the two models that have an active outflow at $t_{SN}$ (HP3 and L1). In these cases, the
progenitor system is surrounded by a small region of unshocked outflow with mass $M_{uof}$ that extends to a radius
$R_{uof}$.

\subsection{Results: Modified Values}
\label{sub:CSM-results-modified}

Among the seven outflow models presented in Table \ref{tab-1}, we have chosen models HP3 and L2 as representative
examples of outflows producing `small' and `large' cavities. Based on these models, we have explored the range of CSM
structures that might be found around Type Ia progenitors by modifying the properties of the ISM and the values of $u_{of}$.

\paragraph{Variation of the ISM properties.} The reference values of $\rho_{ISM}$ and $T_{ISM}$ are at the high end of
the expected ranges for the warm atomic phase of the ISM \citep{ferriere01:ISM}. This choice maximizes the
value of $p_{ISM}$, so that the pressure containment of the wind-blown cavities is maximum, and the size of the
wind-blown bubbles shown in Figure \ref{fig-3} is a lower limit. To evaluate the effect of decreasing $p_{ISM}$ to the
lowest reasonable value for the warm ISM, we have evolved the outflow models HP3 and L2 into an ISM with
$\rho_{ISM}=5\times10^{-25}\,\mathrm{g\,cm^{-3}}$ and $T_{ISM}=5\times10^{3}\,\mathrm{K}$. This translates into an ISM
pressure of $p_{ISM}=3.33\times10^{-13}\,\mathrm{dyn\,cm^{-2}}$, roughly an order of magnitude below the reference
value. The results can be seen in Table \ref{tab-2} and Figure \ref{fig-4}, labeled as models HP3lowp and
L2lowp. These bubbles are $\sim 20 \%$ larger than the reference cases, but there are no significant differences in
the structure of the cavities, because the outflows stay in the fast regime. Even at the lower value of $p_{ISM}$,
model L2lowp is already making the transition to the APCB stage at $t_{SN}$.

\paragraph{Variation of $u_{of}$.} We have calculated the CSM structures obtained by models HP3 and L2 with outflow
velocities of $10\,\mathrm{km\,s^{-1}}$ (a reasonable lower limit, being the typical value for the sound speed in the
ISM), $100\,\mathrm{km\,s^{-1}}$, and $200\,\mathrm{km\,s^{-1}}$. The results of these calculations are presented in
Table \ref{tab-2} and Figure \ref{fig-5} (models Hp3u1e6, Hp3u1e7, Hp3u2e7, L2u1e6, L2u1e7, and L2u2e7). The lowest
values of $u_{of}$ in this sequence (10 and $100\,\mathrm{km\,s^{-1}}$) lead to outflows in the slow regime and
momentum-driven cavities with structures that are very different from the ones we have seen up to now. The types of
cavities described by \citet{koo92:bubbles_I} are nicely laid out by the sequence of HP3 models with increasing values
of $u_{of}$. Model HP3u1e6 is a radiative bubble (RB), with a wind shock that is still radiative at $t_{SN}$; model
HP3u1e7 is a partially radiative bubble (PRB), with an adiabatic wind shock, but active cooling of the shocked outflow
at the CD; model HP3u2e7 is already in the fast regime, and has become pressure confined at $t_{SN}$ (APCB). The
sequence of L2 models reveals an interesting effect associated with accretion winds that have a mass conservative phase
prior to the SN explosion. If these outflows are in the fast regime, the energy-driven cavity survives for a long time,
although it may become pressure confined and start to shrink (c.f. model L2u2e7, which is an APCB at $t_{SN}$). If the
outflow is in the slow regime, on the other hand, the momentum-driven cavities collapse when the accretion wind
ceases. In the case of model L2u1e6, which is a RB during the active outflow phase, a relic shell of cooled material is
left behind, but the PRB of model L2u1e7 (which looked very similar to HP3u1e7 in the active phase) has disappeared
almost completely at $t_{SN}$.

We conclude this section with a brief reference to our previous results from BB01. The hydrodynamic simulations in that
paper were very simplified, and did not include radiative losses. The cavity sizes for the outflows in the fast regime
(models A and C in BB01) are qualitatively correct, but the results for models with outflow velocities of $20\, \mathrm{km
  \, s^{-1}}$ (models B and D), which are in the slow regime, are incorrect \citep[the simulations were later repeated
with the inclusion of radiative losses in][]{badenes04:PhD}. The main conclusion of BB01 - that fast accretion wind
outflows lead to large cavities around Type Ia progenitors - still holds, but the more detailed results that we present
here supersede the calculations in our previous work.

\section{SNR DYNAMICS IN THE MODIFIED CIRCUMSTELLAR MEDIUM}
\label{sec:dynamics-snr}

The dynamics of SNRs expanding into low-density wind-blown cavities has been studied by several authors
\citep{tenorio90:bubblesI,tenorio91:bubblesII,dwarkadas05:SNR-Bubbles_1D}. The reader is encouraged to consult these
works for discussions - we will not go into the details here. Essentially, the SN ejecta expand almost freely until the
inner edge of the cavity $R_{c}$ is encountered. What happens after that depends on the momentum of the ejecta and the
mass that is contained in the shell that surrounds the cavity, $M_{sh}$. \citet{dwarkadas05:SNR-Bubbles_1D} described
this interaction in terms of the $\Lambda$ parameter, defined as the quotient between the shell mass and the ejecta
mass, $\Lambda=M_{sh}/M_{ej}$. Among the CSM structures presented in \S~\ref{sec:CSM}, those generated by outflows in
the fast regime always correspond to the more extreme $\Lambda>>1$ scenario (c.f. values of $M_{sh}$ listed in Table
\ref{tab-2}). In this case, the forward shock (FS) becomes radiative when it hits the shell, and the Sedov stage (i.e.,
$R_{FS} \propto t^{0.4}$) is absent from the SNR evolution \citep[see \S~7.3 in][]{dwarkadas05:SNR-Bubbles_1D}. This
strong modification of the dynamics will have two important observational consequences. First, the fundamental
properties of the FS (radius, velocity, and expansion parameter) at a given SNR age will be very different from the case
of an interaction with a uniform ISM. Second, since most of the ejecta will expand to very low densities before being
overrun by the reverse shock, the X-ray emission from the shocked ejecta in the SNR will also be affected, leading to a much lower
ionization state at a given age. The CSM structures generated by outflows in the slow regime are more diverse, and their
imprint on the FS properties and the X-ray emission from the shocked ejecta will not be so dramatic.

In order to produce SNR models that can be compared with the observations, we have used VH-1 to simulate the interaction
of the Type Ia explosion model PDDe from \citet{badenes03:xray} with several of the CSM profiles presented in
\S~\ref{sec:CSM}. The choice of a particular Type Ia SN explosion model does not have a significant influence on the
radius and velocity of the FS, as long as the kinetic energy produced in the explosion is $\sim 10^{51}$ erg \citep[see
Figs. 3a and 3c in][]{badenes03:xray}. The X-ray emission from the shocked ejecta, on the other hand, is profoundly
affected by the differences in density and chemical composition profiles of Type Ia SN explosion models. We have chosen
model PDDe because it maximizes the ionization state of the elements in the shocked ejecta for a given value of
$\rho_{ISM}$ \citep[compare panel 6d with the other panels of Figure 6 in][]{badenes03:xray}. We have followed the
nonequilibrium ionization (NEI) processes in the shocked ejecta with the ionization code described in
\citet{badenes03:xray}, with one important modification. Some extreme cases of CSM interaction that we will specify in
\S~\ref{sub:x-ray-emission:modvsobs} lead to catastrophic cooling in the outer ejecta layers. To model this process, we
have included ionization and radiative losses in the code, using the latest atomic data published by the CHIANTI
collaboration \citep{dere97:CHIANTI,landi06:CHIANTI}. To avoid recalculating the hydrodynamic evolution in each case, we
have assumed isobaric cooling in the affected layers (i.e., the layers cool but remain in hydrostatic equilibrium with
their surroundings), which is an excellent approximation for the inter-shock region in one-dimensional hydrodynamics.

\section{COMPARISON WITH OBSERVATIONS}
\label{sec:comp-with-obs-SNRs}

\subsection{Observations: Forward Shock Dynamics and X-ray Emission from the Shocked Ejecta}
\label{sub:obs_SNRIa}

We have compiled the relevant observational data for young SNRs that have a firm Type Ia identification in Tables
\ref{tab-3} and \ref{tab-4}. We have restricted ourselves to objects that are close enough to study the dynamics of the
FS, and (if possible) the X-ray emission from the shocked ejecta, and whose age is either known (i.e., historical SNRs)
or can be estimated with a high degree of confidence. This yields seven objects: SN 1885 in M31; Tycho, Kepler and SN
1006 in our Galaxy; and 0509-67.5, 0519-69.0, and N103B in the Large Magellanic Cloud (LMC). Among these objects, SN
1885 was identified as a Type Ia SN from the optical spectrum recorded in the nineteenth century
\citep{vaucoleurs85:SN1885}, although it was probably a subluminous event
\citep{chevalier88:SN1885,fesen06:SN1885}. Tycho and SN 1006 have been traditionally regarded as prototypical Type Ia
SNRs based on their X-ray spectra and other evidence, and their origin has been confirmed by detailed modeling of the
ejecta emission \citep[for Tycho, see][for SN 1006, Badenes et al., in preparation]{badenes05:tycho}. The only object in
our sample whose origin is somewhat controversial is the Kepler SNR \citep[for a discussion and references,
see][]{cassam03:kepler}, but the prominent Fe emission and the virtual absence of O in the shocked ejecta revealed by
the recent deep \textit{Chandra} observations \citep{reynolds06:kepler} are strongly indicative of a Type Ia
origin. The LMC SNRs 0509-67.5, 0519-69.0, and N103B were first identified as Type Ia by
\citet{hughes95:typing_SN_from_SNR}. The results of later works appear to agree with the Type Ia hypothesis in the case
of 0509-67.5 \citep{warren03:0509-67.5} and N103B \citep{lewis03:N103B}, but there are no detailed models in the
literature that can confirm this. An interesting development for these LMC SNRs has been the recent detection of light
echoes from the explosion, which has greatly reduced the uncertainty in their ages and might eventually confirm their
Type Ia origin \citep{rest05:LMC_light_echoes}.

The radius and velocity of the FS in a SNR provide a first assessment of its dynamical state. The FS radii $R_{FS}$ are
determined from the angular radii $\alpha_{FS}$ and the distances to the SNRs $D$ ($R_{FS}=\alpha_{FS} D$). All the
angular radii listed in Table \ref{tab-3} are from \textit{Chandra} observations except that of SN 1885, which is from
the absorption features seen in \textit{HST} images \citep{fesen06:SN1885}. The X-ray radii are very precise: even in
the LMC SNRs, the \textit{Chandra} PSF will only introduce errors below $\sim 5 \%$. The value of $D$ is known with some
accuracy for the LMC SNRs \citep{alves04:LMC_Distance} and for SN 1885 \citep{fesen06:SN1885}, but it can be very
uncertain in the Galactic SNRs. The FS velocities are obtained from the study of the narrow and broad components of the
H$\alpha$ emission in non-radiative shocks \citep{chevalier78:balmerH_nonradiative_shocks}. The derivation of $u_{FS}$
from the width of the broad H$\alpha$ component is not straightforward, and it requires some modeling of the plasma
physics at the shock. It is important to note that the FS velocities are obtained at specific points in the blast wave
(usually those that offer the best signal to noise ratio in H$\alpha$), and may not be representative of the dynamics of
the entire SNR. Furthermore, \citet{heng07:Balmer_Shocks} recently expressed concern that the current models might be
underestimating the shock velocities from the H$\alpha$ observations. For these reasons, the values of $u_{FS}$ listed
in Table \ref{tab-3} need to be considered with caution.

We have used the average ionization timescale $\langle n_{e}t \rangle$ of Fe and Si to represent the ionization state of
these elements in the shocked ejecta of our SNRs. In order to assemble a consistent set of values and constrain the
uncertainties in the best possible way, we have determined the ionization timescales listed in Table \ref{tab-4} from
archival \textit{Chandra} observations of our SNR sample. The methods and techniques that we have used for this purpose
are detailed in Appendix \ref{sec:ioniz-timesc-shock}.

\subsection{Forward Shock Dynamics: Models vs. Observations}
\label{sub:forw-shock-dynam:modvsobs}

Before attempting a comparison between the dynamics of the FS in our SNR models and the observed values, we must
emphasize that our calculations do not include the effect of efficient shock acceleration of cosmic rays. This process
is known to take part in many young SNRs, and recent theoretical \citep{ellison04:hd+cr} and observational
\citep{decourchelle00:cr-thermalxray,warren05:Tycho} works have shown that it can have a noticeable impact on the
internal structure and X-ray emission of the SNR, specially in the shocked ISM. The impact on the dynamics of the FS,
however, is modest: at an age of 5000 yr, the values of $R_{FS}$ and $u_{FS}$ in a SNR that is accelerating cosmic rays
efficiently are only $\sim15\%$ lower than in a test particle case \citep[see Fig. 3 in][]{ellison04:hd+cr}. This kind
of deviation is inconsequential for the order of magnitude comparisons that we carry out in the present Section.

The values of $R_{FS}$ and $u_{FS}$ in our SNR models are plotted alongside the observations in Figures \ref{fig-6} and
\ref{fig-7}. In Figure \ref{fig-6}, we display the values obtained for the interaction between the explosion model PDDe
and the CSM profiles HP3 and L2 from Figure \ref{fig-3}, together with a set of SNR models generated with a uniform
ISM. For the uniform ISM interaction, we have represented the parameter space obtained by varying $\rho_{ISM}$ as a
dashed region between two limiting cases, $\rho_{ISM}=5 \times 10^{-25} \,\mathrm{g\,cm^{-3}}$ (top plots) and
$\rho_{ISM}=5 \times 10^{-24} \,\mathrm{g\,cm^{-3}}$ (bottom plots), which encompass the typical conditions in the warm
atomic phase of the ISM. The SNR models evolving inside the HP3 and L2 CSM profiles are clearly unable to reproduce the
observed values of $R_{FS}$, but most of the data points cluster nicely around the uniform ISM models. The only
exception is SN 1006, whose large size is compatible with both a very low density ISM and the CSM cavity from model
HP3. The situation is somewhat different for the FS velocities. While it is clear that the SNR models evolving inside
the CSM cavities can be confidently discarded, it seems that the models interacting with a uniform ISM also overpredict
the values of $u_{FS}$, again with the only exception of SN 1006. The distribution of the data points for Kepler, Tycho,
0509-67.5 and 0519-69.0 suggests a systematic shift of a factor $\sim\,2$ between the values of $u_{FS}$ predicted by
modeling the H$\alpha$ emission of the FS and the models with an ISM interaction. This could be the hallmark of the
effect discussed by \citet{heng07:Balmer_Shocks}, or a selection effect towards localizations in the FS with higher
densities imposed by the statistics of the H$\alpha$ observations, but more complex explanations cannot be discarded
without a detailed analysis.

In Figure \ref{fig-7}, we repeat the comparison, this time for SNRs evolving into the cavities obtained with the
modified values of $u_{of}$. From the point of view of the FS dynamics alone, the CSM structures produced by outflows in
the slow regime are hard to distinguish from a uniform ISM interaction. This is specially true for the collapsed
cavities of models L2u1e6 and L2u1e7 and the small RB of model HP3u1e6. The energy-driven cavities of models HP3u2e7 and
L2u2e7, on the other hand, are hard to reconcile with the observations of all SNRs except SN 1006. The size of these two
cavities is coincidentally similar to the radius of SN 1006, so that the FS is hitting the shell at $R_{c}$ for a SNR
age of $\sim\,1000$ yr. This results in a good match to the present value of $u_{FS}$ in SN 1006, but in a regime where
the FS is undergoing a very rapid deceleration as it overcomes the radiatively cooled shell (dash-dotted plots in Figure
\ref{fig-7}). If this was indeed the situation, one would expect to see a radiative FS with a very low expansion
parameter in SN 1006, instead of the observed nonradiative FS \citep{winkler03:SN1006_Distance} with an expansion
parameter of $\sim \, 0.48$ \citep{moffett93:Expansion_SN1006_Radio}.

We conclude this section noting that there is evidence for some kind of CSM interaction in at least two of the objects
in our sample, Kepler \citep{bandiera87:kepler,borkowski92:kepler} and N103B \citep{lewis03:N103B}. The fact that this
is not clearly revealed by the values of $R_{FS}$ and $u_{FS}$ listed in Table \ref{tab-3} emphasizes the importance
that hydrodynamic models of individual SNRs will have in order to take these comparisons to the next level of detail.

\subsection{X-ray Emission from the Shocked Ejecta: Models vs. Observations}
\label{sub:x-ray-emission:modvsobs}

In Figures \ref{fig-8} and \ref{fig-9}, we compare the values of $\langle n_{e}t \rangle$ for Fe and Si listed in Table
\ref{tab-4} to the emission measure averaged ionization timescales \citep[$\langle \tau \rangle$ as defined in \S~4.2
of][]{badenes03:xray} in our models. These comparisons have to be of a more qualitative sort than those based on the FS
dynamics, both due to issues that affect the spectral models that we use to fit the X-ray observations (see Appendix
\ref{sec:ioniz-timesc-shock}) and to the limitations of our one-dimensional hydrodynamic calculations \citep[see
discussion in \S~8 of][]{badenes05:tycho}. In spite of this, we find that qualitative comparisons of ionization
timescales are highly informative, and more than sufficient for our present needs.

In Figure \ref{fig-8}, we present the temporal evolution of $\langle \tau \rangle$ for Si and Fe in the SNR models
obtained from the interaction of model PDDe with the CSM profiles HP3 and L2, and with a uniform ISM. For the uniform
ISM case, we have represented the variation of $\rho_{ISM}$ between $\rho_{ISM}=5 \times 10^{-25} \,\mathrm{g\,cm^{-3}}$
and $\rho_{ISM}=5 \times 10^{-24} \,\mathrm{g\,cm^{-3}}$ as in Figures \ref{fig-6} and \ref{fig-7}. We have scaled
$\langle \tau \rangle$ with $\rho_{ISM}$ using the approximate relations given in \S~2.3 of \citet{badenes05:xray}. Once
again, the data points cluster around the models with a uniform ISM interaction, albeit with more dispersion than in the
$R_{FS}$ plots. The inconsistency between the SNR models evolving inside low-density cavities and the observations is
even more dramatic than for the FS dynamics. There is simply not enough mass in the CSM models HP3 and L2 that the SN
ejecta can react to in order to reach the ionization timescales observed in Type Ia SNRs.

In Figure \ref{fig-9}, we repeat the comparison using the CSM profiles obtained with the modified values of $u_{of}$. As
expected, the CSM profiles generated by outflows in the slow regime lead to a stronger interaction with the SN ejecta
and higher ionization timescales for Fe and Si. This is specially true for models HP3u1e6 and L2u1e6, which have
prominent shells of radiatively cooled material close to the SN progenitor. In these models, some ejecta layers undergo
catastrophic cooling early on the evolution of the SNR. Models HP3u1e7 and L2u1e7 do not lead to such extreme
interaction, and are in general hard to distinguish from a uniform ISM. As soon as the outflows enter the fast regime,
however, the situation changes radically. Only the shocked ejecta emission of SN 1006, with its low Si $n_{e}t$ and
absence of Fe, is compatible with the CSM cavities HP3u2e7 and L2u2e7, but these models have already been discarded in
\S~\ref{sub:forw-shock-dynam:modvsobs} for other reasons.

\section{DISCUSSION}
\label{sec:Discussion}

In \S~\ref{sec:CSM}, we have seen that the accretion wind outflows invoked by current Type Ia progenitor models in the
SD channel will excavate large cavities in the ISM. In \S~\ref{sec:comp-with-obs-SNRs}, we have shown that the
fundamental properties of the seven Type Ia SNRs in our sample are incompatible with SNR models that expand into such
large cavities. Can these two results be connected with enough confidence to draw conclusions on Type Ia SN progenitor
models based on the SNR observations? In this Section, we review the potential issues that might affect the initial
models, the simulations, and the observational sample, and we briefly discuss the implications of our results for the
observations of Type Ia SNe.

\subsection{Initial Models: The True Nature of Accretion Winds.}
\label{sub:Discuss-initial_models}

\begin{figure}

  \centering
 
  \includegraphics[angle=90,scale=0.75]{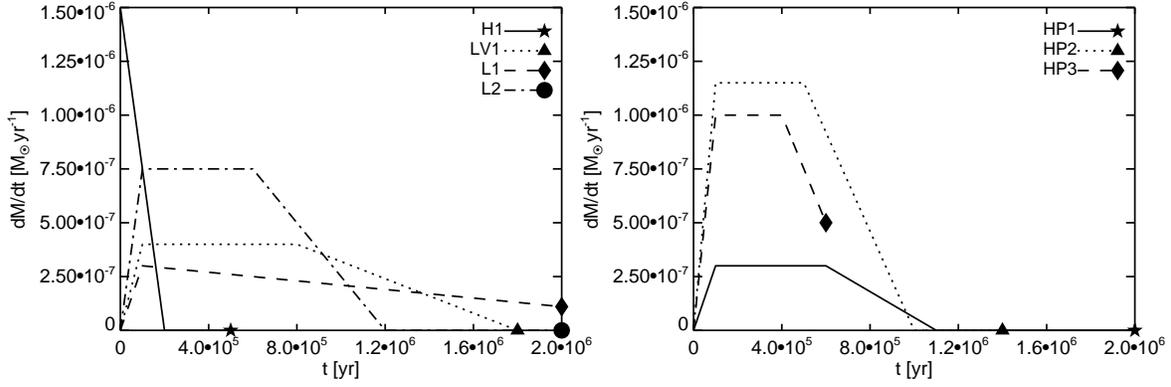}

  \caption{Temporal evolution of the mass loss rates in the outflow models for Type Ia SN progenitor systems detailed in
    Table \ref{tab-1}. Left: models H1, LV1, L1, and L2. Right: models HP1, HP2, and HP3. The symbols mark the time of
    the SN explosion, $t_{SN}$. \label{fig-1}}

\end{figure}

\begin{figure}

  \centering
 
  \includegraphics[angle=90,scale=0.75]{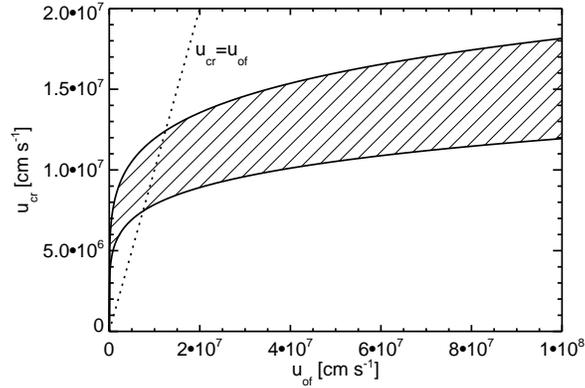}

  \caption{Critical outflow velocity $u_{cr}$ as a function of outflow velocity $u_{of}$. The striped region represents
    the values of $u_{cr}$ for typical accretion wind outflows expanding into the warm, atomic phase of the ISM. The
    limiting curves are for $\dot{M}_{of} = 10^{-6} \,\mathrm{M_{\odot}\, yr^{-1}}$, $\rho_{ISM}=5 \times
    10^{-24}\,\mathrm{g\,cm^{-3}}$ (top curve) and $\dot{M}_{of} = 10^{-7} \,\mathrm{M_{\odot} \, yr^{-1}}$,
    $\rho_{ISM}=5 \times 10^{-25}\,\mathrm{g\,cm^{-3}}$ (bottom curve). The dotted line has a slope of 1.\label{fig-2}}

\end{figure}

\begin{figure}

  \centering
 
  \includegraphics[angle=90,scale=0.75]{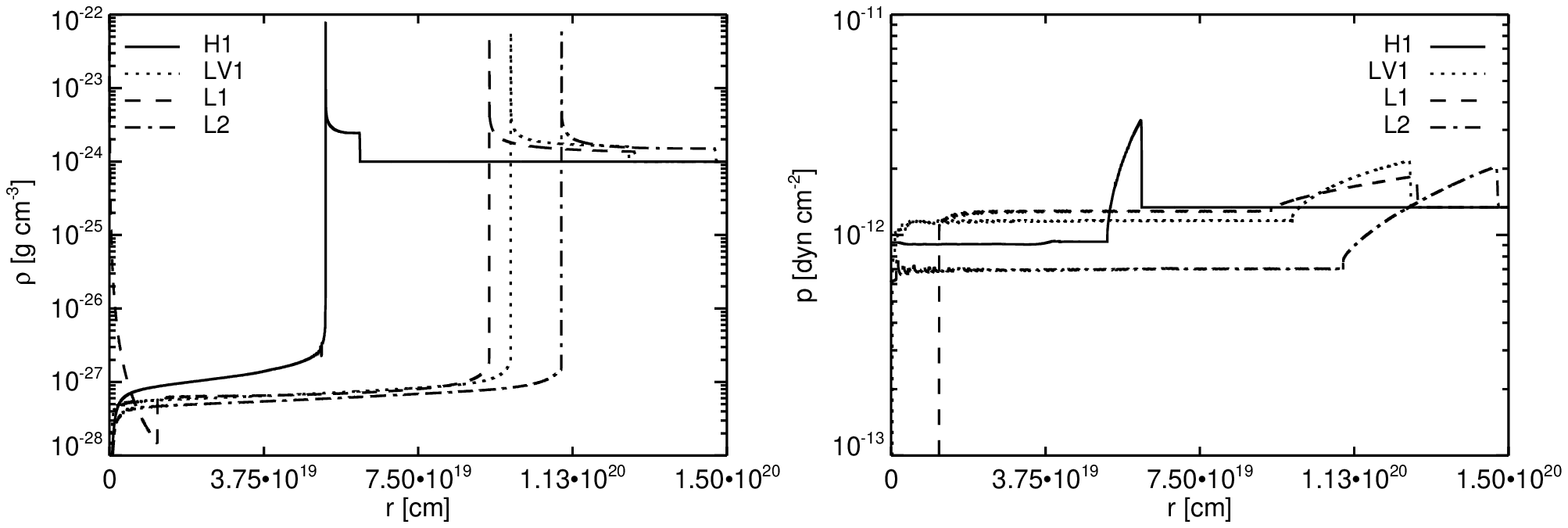}
  \includegraphics[angle=90,scale=0.75]{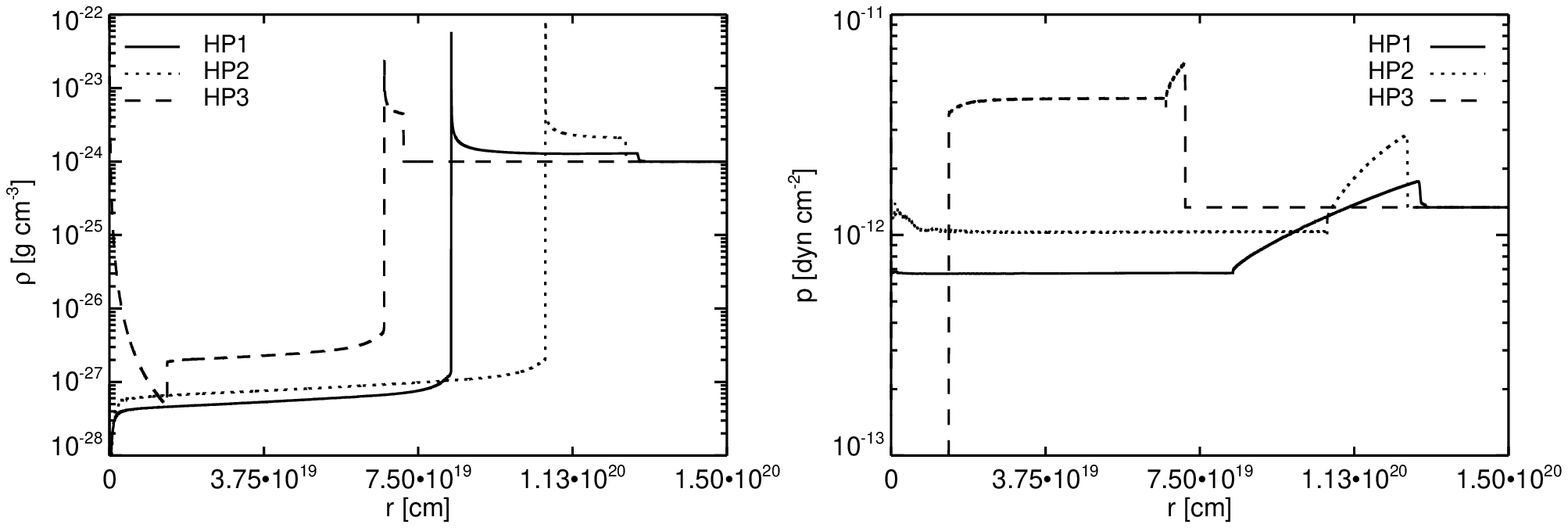}

  \caption{Radial structure of the CSM sculpted by the outflow models with the reference values:
    $u_{of}=10^{3}\,\mathrm{km \, s^{-1}}$, $\rho_{ISM}=10^{-24}\,\mathrm{g \,cm^{-3}}$, and
    $T_{ISM}=10^{4}\,\mathrm{K}$. Density is on the left panels and pressure is on the right panels.\label{fig-3}}

\end{figure}

\begin{figure}

  \centering
 
  \includegraphics[angle=90,scale=0.75]{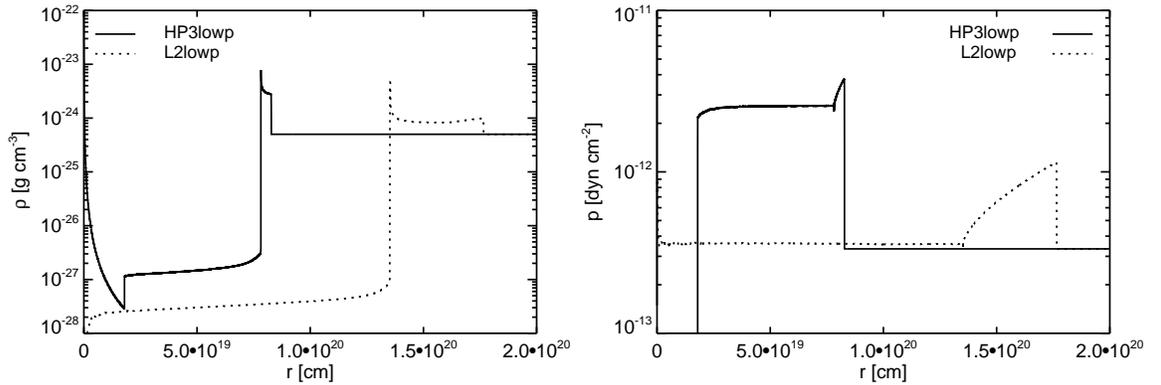}

  \caption{Radial structure of the CSM sculpted by the outflow models HP3 and L2 with
    $\rho_{ISM}=5\times10^{-25}\,\mathrm{g \,cm^{-3}}$, and $T_{ISM}=5\times10^{3}\,\mathrm{K}$. Density is on the left
    panel and pressure is on the right panel.\label{fig-4}}

\end{figure}

\begin{figure}

  \centering
 
  \includegraphics[angle=90,scale=0.75]{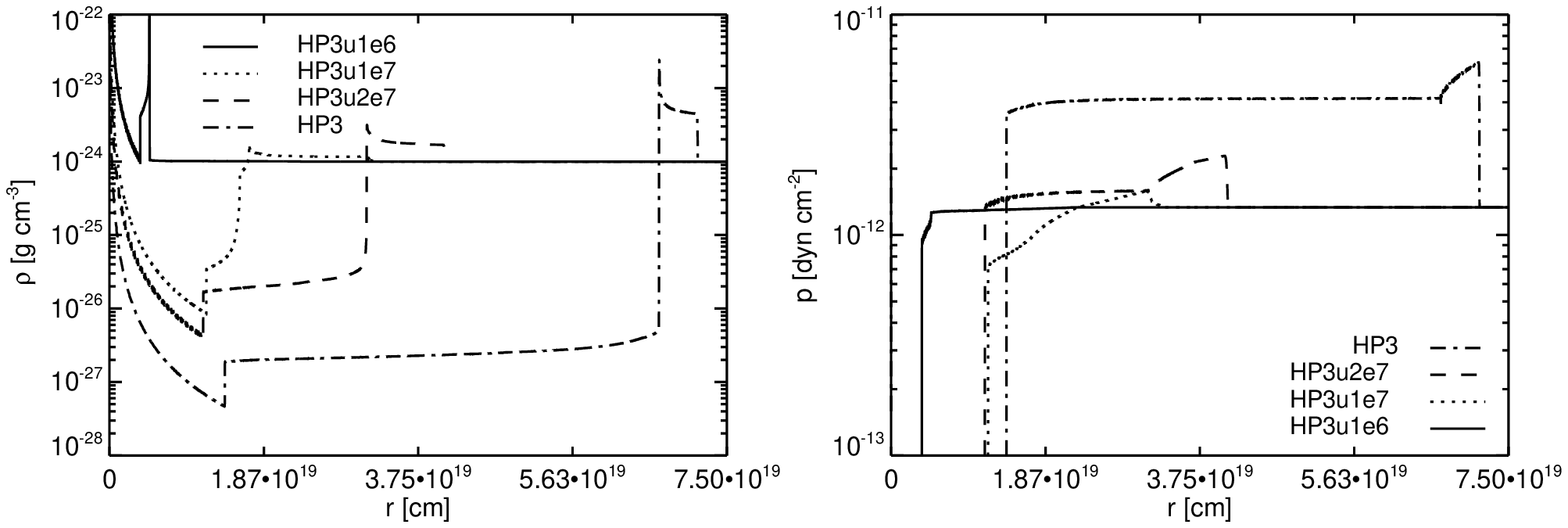}
  \includegraphics[angle=90,scale=0.75]{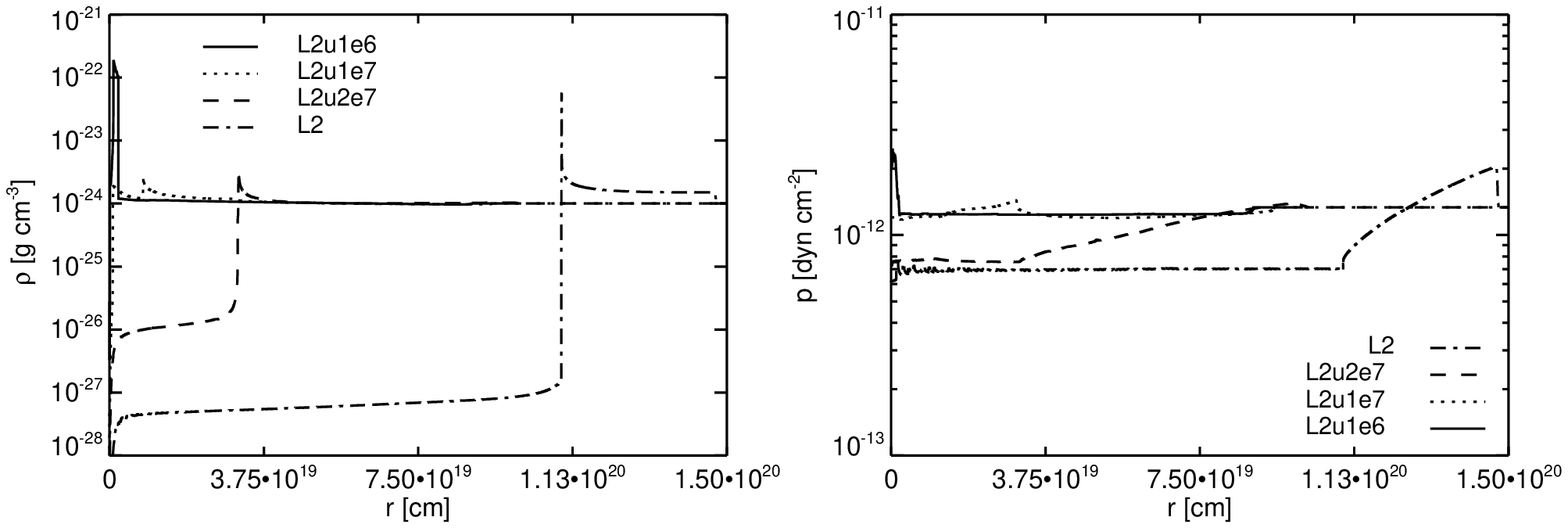}

  \caption{Radial structure of the CSM sculpted by the outflow models HP3 (top panels) and L2 (bottom panels), with
    $u_{of}=10\,\mathrm{km \, s^{-1}}$ (solid plots), $u_{of}=10^{2}\,\mathrm{km \, s^{-1}}$ (dotted plots),
    $u_{of}=2\times10^{2}\,\mathrm{km \, s^{-1}}$ (dashed plots), and $u_{of}=10^{3}\,\mathrm{km \, s^{-1}}$
    (dash-dotted plots - the same as in Figure \ref{fig-3}, added for comparison). Density is on the left panels and
    pressure is on the right panels.\label{fig-5}}

\end{figure}

\begin{figure}

  \centering
 
  \includegraphics[angle=90,scale=0.75]{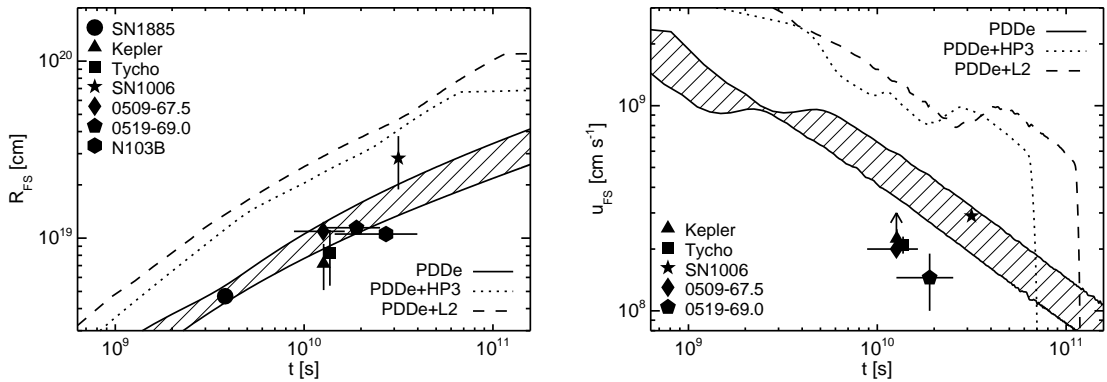}

  \caption{Temporal evolution of the forward shock radius $R_{FS}$ (left panel) and forward shock velocity $u_{FS}$
    (right panel) for Type Ia SN explosion model PDDe interacting with a constant density ISM (solid lines), and with the
    CSM profiles HP3 (dotted line) and L2 (dashed line). The two solid plots correspond to an interaction with
    $\rho_{ISM}=5 \times 10^{-25}\,\mathrm{g\,cm^{-3}}$ (top plots) and $\rho_{ISM}=5 \times
    10^{-24}\,\mathrm{g \, cm^{-3}}$ (bottom plots). The striped region between the solid lines spans the parameter space
    of $\rho_{ISM}$ between these two values. The data from Table \ref{tab-4} are overlaid on the plots.\label{fig-6}}

\end{figure}

\begin{figure}

  \centering
 
  \includegraphics[angle=90,scale=0.75]{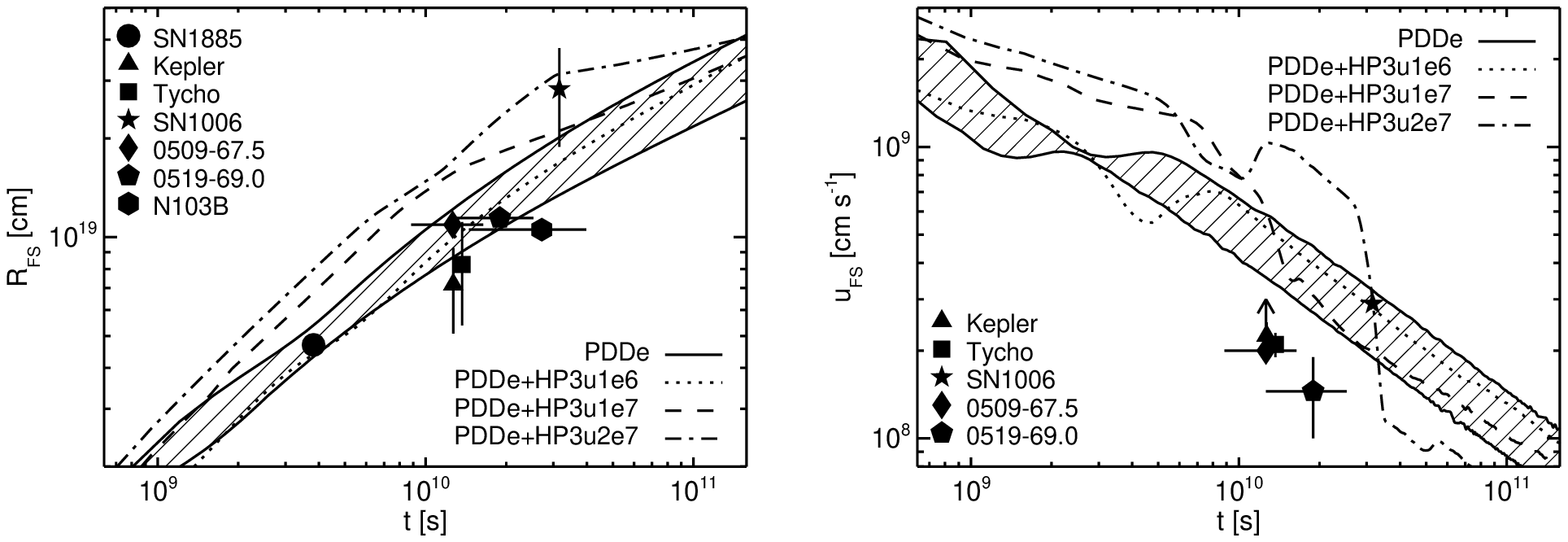}
  \includegraphics[angle=90,scale=0.75]{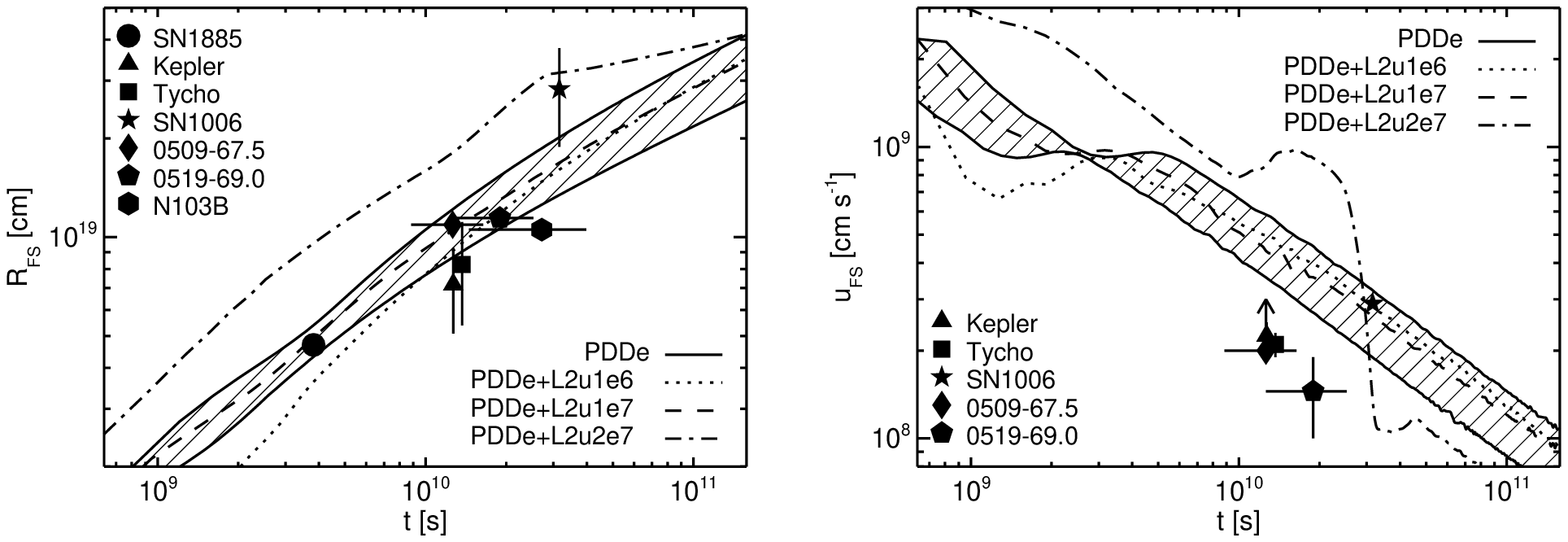}

  \caption{The same as Figure \ref{fig-6}, but for the explosion model PDDe interacting with a uniform ISM (solid
    lines), and the CSM profiles generated by outflows with modified velocities. Top panels: models HP3u1e6 (dotted
    line), HP3u1e7 (dashed line), HP3u2e7 (dash-dotted line). Bottom panels: models L2u1e6 (dotted line), L2u1e7 (dashed
    line), L2u2e7 (dash-dotted line).\label{fig-7}}

\end{figure}

\begin{figure}

  \centering
 
  \includegraphics[angle=90,scale=0.75]{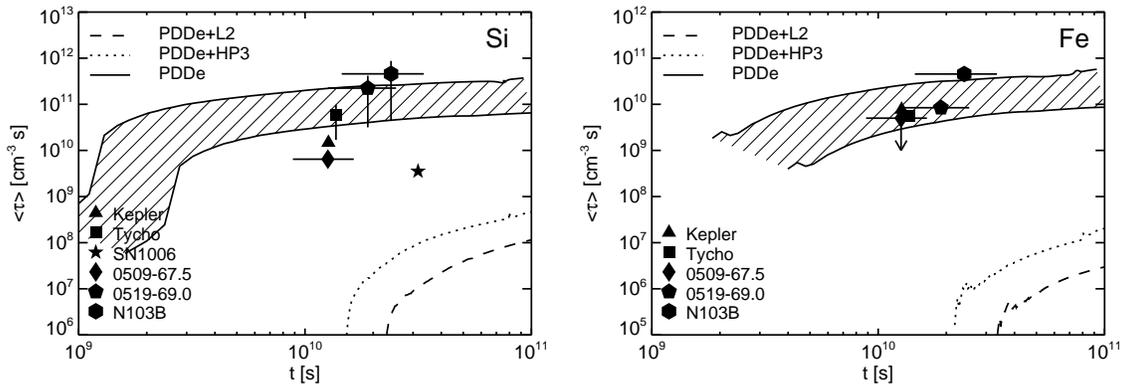}

  \caption{Temporal evolution of the emission measure averaged ionization timescale $\langle \tau \rangle$ for Si (left
    panel) and Fe (right panel) in the shocked ejecta of model PDDe interacting with a constant density ISM (solid
    lines), and with the CSM profiles HP3 (dotted line) and L2 (dashed line). The two solid plots correspond to an
    interaction with $\rho_{ISM}=5 \times 10^{-25}\,\mathrm{g\,cm^{-3}}$ (top plots) and $\rho_{ISM}=5 \times
    10^{-24}\,\mathrm{g \, cm^{-3}}$ (bottom plots). The striped region between the solid lines spans the parameter
    space of $\rho_{ISM}$ between these two values. The data from Table \ref{tab-4} are overlaid on the
    plots.\label{fig-8}}

\end{figure}

\begin{figure}

  \centering
 
  \includegraphics[angle=90,scale=0.75]{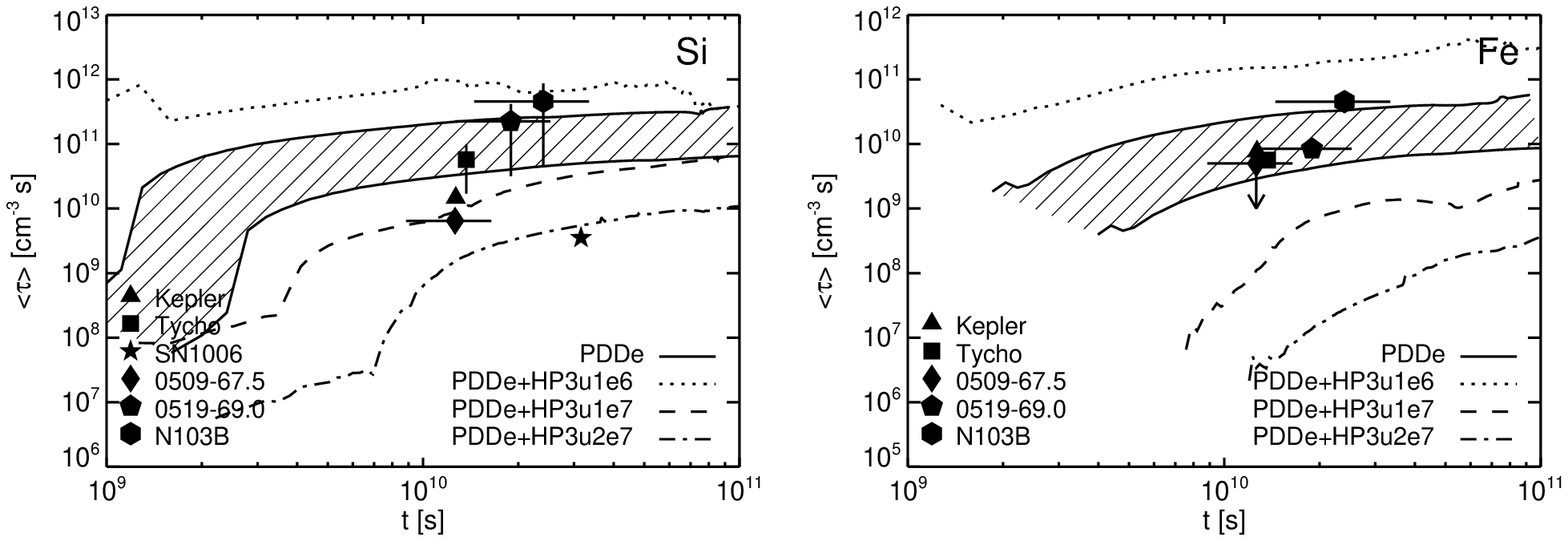}
  \includegraphics[angle=90,scale=0.75]{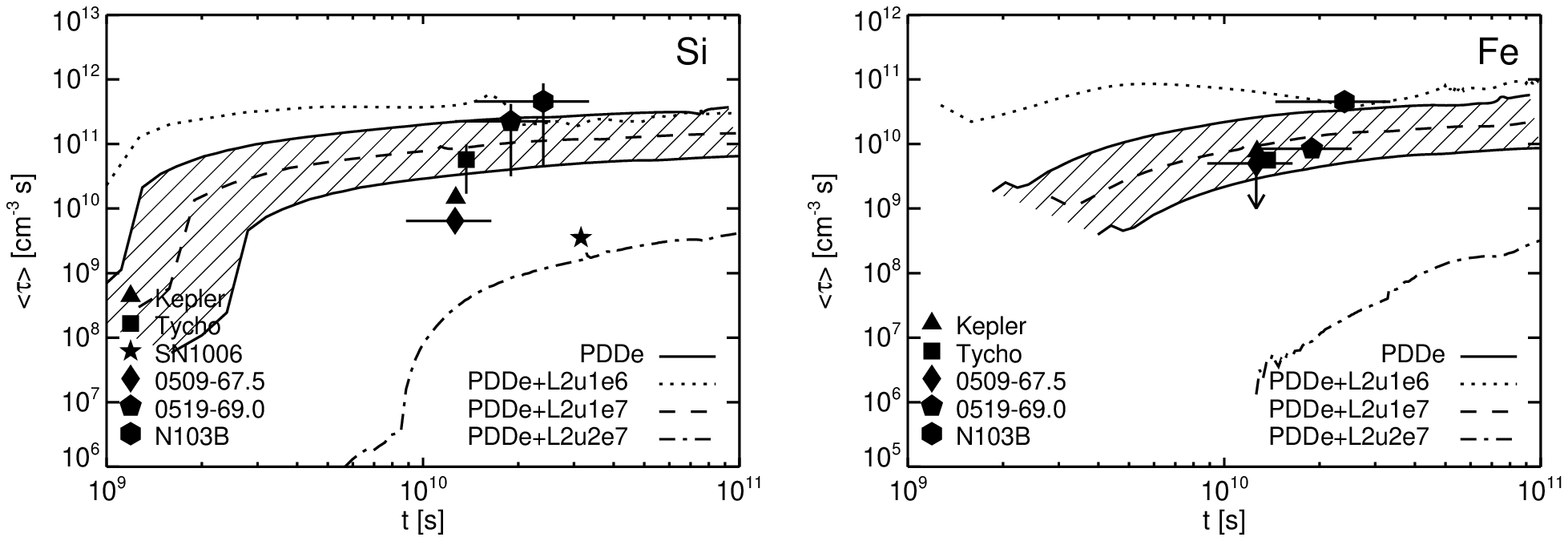}

  \caption{The same as Figure \ref{fig-8}, but for the explosion model PDDe interacting with a uniform ISM (solid
    lines), and the CSM profiles generated by outflows with modified velocities. Top panels: models HP3u1e6 (dotted
    line), HP3u1e7 (dashed line), HP3u2e7 (dash-dotted line). Bottom panels: models L2u1e6 (dotted line), L2u1e7 (dashed
    line), L2u2e7 (dash-dotted line). \label{fig-9}}

\end{figure}

The outflow models presented in Figure \ref{fig-1} and Table \ref{tab-1} make two important approximations: they are
isotropic (i.e., spherically symmetric) in space and continuous in time. In a more realistic scenario, the presence of
the donor star and the accretion disk, and the spin-up of the WD to high angular velocities
\citep{yoon04:no_subCh,yoon05:rotating_WD_SNe_Collapse} should introduce some sort of bipolarity in the outflows. As we
have seen in \S~\ref{sub:outflows-observations}, the observations of Type Ia SN progenitor candidates show some evidence
of this bipolarity. Continuous bipolar outflows may lead to a bipolar CSM structure similar to some planetary nebulae
\citep{balick02:PNe}, but this does not seem likely for the objects in our sample. Only moderate one-sided asymmetries
are apparent in Kepler and N103B, and most other objects (in particular, Tycho, SN 1006, and 0509-67.5) are nearly
spherical.

Setting aside the spatial structure of the outflows from Type Ia SN progenitors, their temporal evolution will not be as
simple as the plots shown in Figure \ref{fig-1}. In order to reach the values of $\dot{M}$ that allow for the growth of
the WD, the mass transfer rate has to cross the instability region below $\dot{M}_{stable}$, where more or less frequent
nova outbursts are expected. \citet{langer00:prog-models}, for instance, found a long ($\sim 10^{6}\,\mathrm{yr}$)
switch-on phase of the mass transfer in all their binary systems that was dominated by instabilities. Nova-like
outbursts could also appear during the mass-conservative phase between the cessation of the accretion wind and the SN
explosion \citep{hachisu99:supersoft}. Judging from the observations of RXJ0513.9-6951 and V Sagittae discussed in
\S~\ref{sub:outflows-observations}, a steady accretion wind might be impossible or happen only in a few cases. If the
episodic outflows in Type Ia progenitors have periodicities of $\sim 100$ days, as in these two systems, several million
cycles should take place through the accretion phase, and the outflows might smear out to become indistinguishable from
a continuous wind. In the case of recurrent novae, the individual outbursts will be more spaced out, leading to a
complex CSM structure, with several shells that brighten up as they are overtaken by the FS. Again, there is no evidence
for this in any of the objects that we have studied, but more detailed hydrodynamic calculations would be necessary to
confirm or discard any given scenario.

\subsection{Simulations: From Outflows to Cavities.}
\label{sub:Discuss-sim}

There are two ways to prevent Type Ia progenitor outflows from leaving large cavities in the CSM: either the value of
$u_{of}$ is decreased below the critical limit $u_{cr}$ or the cavities are somehow destroyed before the SN explodes. As
we have shown in \S~\ref{sec:comp-with-obs-SNRs}, outflows in the slow regime ($u_{of} \lesssim
100\,\mathrm{km\,s^{-1}}$) lead to CSM structures that are hard to distinguish from a uniform ISM interaction. However,
outflows emanating from the WD surface \textit{should} have much higher velocities, of the order of the escape velocity
($\sim 1000\,\mathrm{km\,s^{-1}}$, a value that is supported by the observations cited in
\S~\ref{sub:outflows-observations}). Some of the kinetic energy of the outflows will be lost in overcoming the potential
well of the system, but it is unlikely that this can reduce $u_{of}$ by an order of magnitude. Furthermore, the presence
of slow outflows around the progenitor systems is difficult to reconcile with the lack of CSM detections in Type Ia SNe
(see \S~\ref{sub:Discuss-comp-with-obs-SNe}).

If large cavities do form, it is hard to find a non-catastrophic process that could destroy them before
$t_{SN}$. Thermal conduction, for instance, which must be active to some extent at the interface between the hot shocked
outflow and the radiatively cooled shell, could not wipe out cavities with a radius of $\sim 10^{20}$ cm, even over
timescales of $\sim 10^{6}$ yr. In the example presented in \S~IV of \citet{weaver77:interstellar_bubbles_II}, which is
based on an interstellar bubble very similar to our models, the conduction front behind the cool shell only spans a
small fraction of the cavity radius. The density inside the cavity increases due to evaporation of material from the
shell, but it remains orders of magnitude below the surrounding ISM.

A final possibility is that the binary system forms a cavity, and then leaves it behind as it moves with respect to the
ISM. Large speeds are not expected in SD Type Ia progenitors, but even at the mean random velocity of
$\sim\,20\,\mathrm{km\,s^{-1}}$ \citep[characteristic of the solar neighborhood,][]{dehnen98:local_kinematics_HIPPARCOS}
a star can move $\sim 10^{20}$ cm in $2 \times 10^{6}$ yr. Still, only moving systems with an extended mass conservative
phase before the SN explosion could leave their cavities completely behind. In moving progenitors with active outflows
at $t_{SN}$, the explosion would happen off-center, but the cavity would still be there, and therefore the impact on the
FS dynamics and the X-ray emission of the SNR would still be strong
\citep{weaver77:interstellar_bubbles_II,rozyczka93:bubblesIII}.

\subsection{Observations: Other Type Ia SNRs?}
\label{sub:Discuss-obs}

We have omitted from our study several SNRs whose X-ray spectra suggest a Type Ia origin, but whose ages can only be
roughly estimated from the FS dynamics. Examples include G337.2-0.7 and G299.2-2.9 in the Galaxy
\citep{rakowski05:G337,park06:G299}, and DEM L71, DEM L238, and DEM L249 in the LMC
\citep{hughes03:DEML71,borkowski06:DEML238_DEML249}. These are rather old objects, and nothing in their morphology or
X-ray emission indicates that they might be expanding inside a large wind-blown cavity, but clues for other kinds of CSM
interaction might be revealed by detailed models.

It is also possible that young Type Ia SNRs expanding inside large cavities \textit{do} exist, but have not been
identified yet. These objects would have a large radius and a faint X-ray emission, so they would be hard to classify as
Type Ia SNRs. One such object might be the Galactic SNR RCW 86, which has a large size ($\alpha_{FS}=42'$) and Fe-rich
SN ejecta \citep{vink97:RCW86,rho02:RCW86}. This SNR has been associated with the supernova of AD 185
\citep{vink06:RCW86}, but this association remains controversial, and its Type Ia origin is uncertain. If we ignore
these concerns, and take the value of $D=2.8$ kpc given by \citet{rosado96:kinematics_RCW86}, we find $R_{FS}=1.1 \times
10^{20}\,\mathrm{cm}$ at an age of 1821 yr, about 30 $\%$ larger than model PDDe+L2, but certainly within the correct
order of magnitude. Recent values for the FS velocity \citep[$\sim 2700\,\mathrm{km\,s^{-1}}$,][]{vink06:RCW86} and the
ionization timescale of the shocked ejecta \citep[$\sim 10^{9}\,\mathrm{cm^{-3}\,s}$,][]{rho02:RCW86} are much closer to
our cavity models PDDe+HP3 and PDDe+L2 than those of the SNRs in our sample. In fact, \citet{vink97:RCW86} already
suggested that RCW 86 might be evolving inside a low-density cavity based on early \textit{ASCA} observations. If the
explosion type and the association with SN 185 are confirmed for RCW 86, it would be very interesting to explore the
possibility of an accretion wind outflow as the origin of the CSM structure. Another Galactic SNR with Fe-rich ejecta
that might be expanding into a CSM cavity is W49B \citep{hwang00:W49B,keohane07:W49B}, but in this case there is no
reliable age estimate that can be used to constrain the dynamics.

\subsection{Implications of accretion wind outflows for the observations of Type Ia SNe}
\label{sub:Discuss-comp-with-obs-SNe}

Although this paper focuses on the impact that the outflows from Type Ia progenitors have on the dynamics and X-ray
emission of SNRs, here we will briefly discuss the implications for observations of Type Ia SNe. The lack of prompt
emission at X-ray and radio wavelengths \citep[][and other references listed in
\S~\ref{sec:Intro}]{panagia06:radio_SNIa,immler06:Swift_Ia_SN2005ke}, as well as the absence of low velocity (narrow)
lines from H in the early or late spectra of normal events \citep{mattila05:noH_in_SN2001el} suggest that the amount of
circumstellar material in the immediate vicinity of Type Ia progenitors must be low
\citep{eck02:radio_emission_SNe}. The upper limits on the radio and X-ray fluxes have been used to derive estimates for
the mass loss rate of the progenitor system outflows $\dot{M}_{of}$, but these observations can only constrain the
quotient $\dot{M}_{of} / u_{of}$. If the outflows are fast, this technique does not provide much information about
$\dot{M}_{of}$: the radio fluxes of \citet{panagia06:radio_SNIa} translate into $\dot{M}_{of} \lesssim 3 \times 10^{-6}
\mathrm{M_{\odot}\,yr^{-1}}$ for $u_{of}=1000\,\mathrm{km\,s^{-1}}$, which is compatible with all but the most extreme
of accretion winds. For slower outflows, the upper limits on $\dot{M}_{of}$ go down linearly with $u_{of}$. An
interesting possibility, suggested by \citet{wood-vasey06:novae_cavities}, is that episodic outflows just before the
explosion might clear out a more or less extended region around the progenitor system, thus explaining the lack of
prompt emission and H-rich material. However, we have shown that this region cannot be so large that the dynamic
evolution of the SNR is severely affected, at least in the seven objects that we have studied. At the same time,
evidence for large CSM structures around \textit{some} Type Ia progenitors has appeared in the form of the light echoes
detected in three objects: SN 1991T, SN1995E and SN 1998bu. These echoes have been interpreted by
\citet{quinn06:SN1995E_light_echo} as reflected light from detached shells or sheets of dust at distances ranging
between 50 and 200 pc from the SN. If this interpretation is correct \citep[see][for a complete discussion of the
ambiguities involved]{patat05:light_echoes_1}, the detached structures in these three objects might be the radiatively
cooled shells around accretion wind-blown bubbles.

\section{CONCLUSIONS}
\label{sec:Conclusions}

In this paper, we have explored the relationship between the SD progenitor systems of Type Ia SNe and their
surroundings, focusing on the effects that the outflows from these systems would have on the structure of the CSM and on
the dynamics and X-ray emission of the SNRs that evolve after the explosion. We have seen that the optically thick
accretion winds from the WD surface invoked by current Type Ia progenitor models will excavate large cavities in the
ISM, provided that the outflows stay in the fast regime ($u_{of} \gtrsim 200\,\mathrm{km\,s^{-1}}$). We have shown that
the fundamental properties of the seven young Type Ia SNRs in our sample (SN 1885, Kepler, Tycho, SN 1006, 0509-67.5,
0519-69.0, and N103B) are incompatible with SNR models that expand into such large cavities. In fact, we found that all
these objects can be explained by a uniform ISM interaction at the level of detail allowed by our simulations. At the
same time, we cannot discard the existence of a population of young Type Ia SNRs expanding into low-density cavities
excavated by accretion wind outflows. The properties of these objects (large size and low brightness) would make them
hard to detect and, if detected, very difficult to identify as the remnants of Type Ia explosions. The Galactic SNR RCW
86 might be such an object, but more detailed work is needed to confirm this intriguing possibility.


The presence of outflows from the progenitors of the seven SNRs that we have examined cannot be completely discarded,
but the properties of these outflows (mass-loss rates, timescales and velocities) must be very different from those of
the accretion winds described by HKN96. It remains to be seen whether outflows with substantially lower values of
$\dot{M}_{of}$ or $u_{of}$ would be capable of limiting the effective accretion rate of the WD and prevent a common
envelope phase in single degenerate Type Ia progenitors. Another possibility is that accretion wind outflows are not
always present in these systems - for instance, if the rotation of the WD increases the accretion efficiency to the
point that only a small fraction of systems present significant outflows, as proposed by
\citet{yoon05:rotating_WD_SNe_Collapse}.

We conclude with a reminder that the viability of SD systems as Type Ia progenitors has not been proved yet. Several
studies have pointed out the difficulties in driving a H-accreting WD close enough to the Chandrasekhar mass
\citep{cassisi98:H-accreting-CO-WD,piersanti99:H_accreting_CO_WD_I,piersanti00:H_accreting_CO_WD_II}. If the star found
by \citet{ruiz-lapuente04:Tycho_Binary} in the Tycho SNR is indeed the runaway companion of the WD that exploded in
1572, this means that at least one of the objects in our sample must have had a SD progenitor. The obvious alternative
(DD systems) is plagued by its own problems, and it is still unclear whether WD mergers can lead to healthy explosions
\citep{segretain97:_fate_mergin_white_dwarf,guerrero04:WD_mergers}. The fact remains that we lack a clear picture of the
binary evolution leading to Type Ia SNe. Studies of the CSM structure in young Type Ia SNRs should provide valuable
insights into this problem.

\acknowledgements We are very grateful to John Blondin for his assistance in setting up the VH-1 code, and for many
discussions and valuable insights regarding the hydrodynamic simulations presented here. Kazik Borkowski provided us
with several suggestions, including the importance of stellar motions and the relevance of RCW 86. We also acknowledge
fruitful discussions on diverse topics with Martin Laming, Steve Reynolds, Jeno Sokoloski, Craig Wheeler, Philipp
Podsiadlowski, and Garrelt Mellema. Support for this work was provided by the National Aeronautics and Space
Administration through Chandra Postdoctoral Fellowship Award Number PF6-70046 issued by the Chandra X-ray Observatory
Center, which is operated by the Smithsonian Astrophysical Observatory for and on behalf of the National Aeronautics and
Space Administration under contract NAS8-03060. JPH acknowledges support from \textit{Chandra} grant G06-7016B. EB has
received support from the DURSI of the Generalitat de Catalunya and the Spanish DGICYT grants AYA 2004-06290-C02-02 and
AYA 2005-08013-C03-01.

\appendix
\section{IONIZATION TIMESCALES FOR THE SHOCKED EJECTA IN TYPE Ia SUPERNOVA REMNANTS}
\label{sec:ioniz-timesc-shock}

Together with the electron temperature $T_{e}$, the ionization timescale or fluence ($n_{e}t$) is a fundamental quantity
used to characterize the thermal emission in nonequilibrium ionization (NEI) plasmas. These two parameters are the most
common end product of fitting the thermal component of the X-ray emission in SNRs. However, the values of $T_{e}$ and
$n_{e}t$ obtained in a particular spectral fit are model dependent, and can be affected by uncertainties in the atomic
data and other factors \citep[for discussions, see][]{borkowski01:sedov,rakowski05:G337}. The most popular tools for
spectral analysis of the thermal emission in SNRs are the plane-parallel shock models with adjustable abundances, but
even these sophisticated spectral models contain fundamental simplifications whose impact on the fitted parameters is
unclear, like the assumption of a homogeneous chemical composition or the plane shock geometry. In general, it can be
said that the systematic uncertainties associated with plane-parallel shock fits are much larger (and much harder to
estimate) than the statistical ones.

\begin{figure}

  \centering
 
  \includegraphics[scale=0.4]{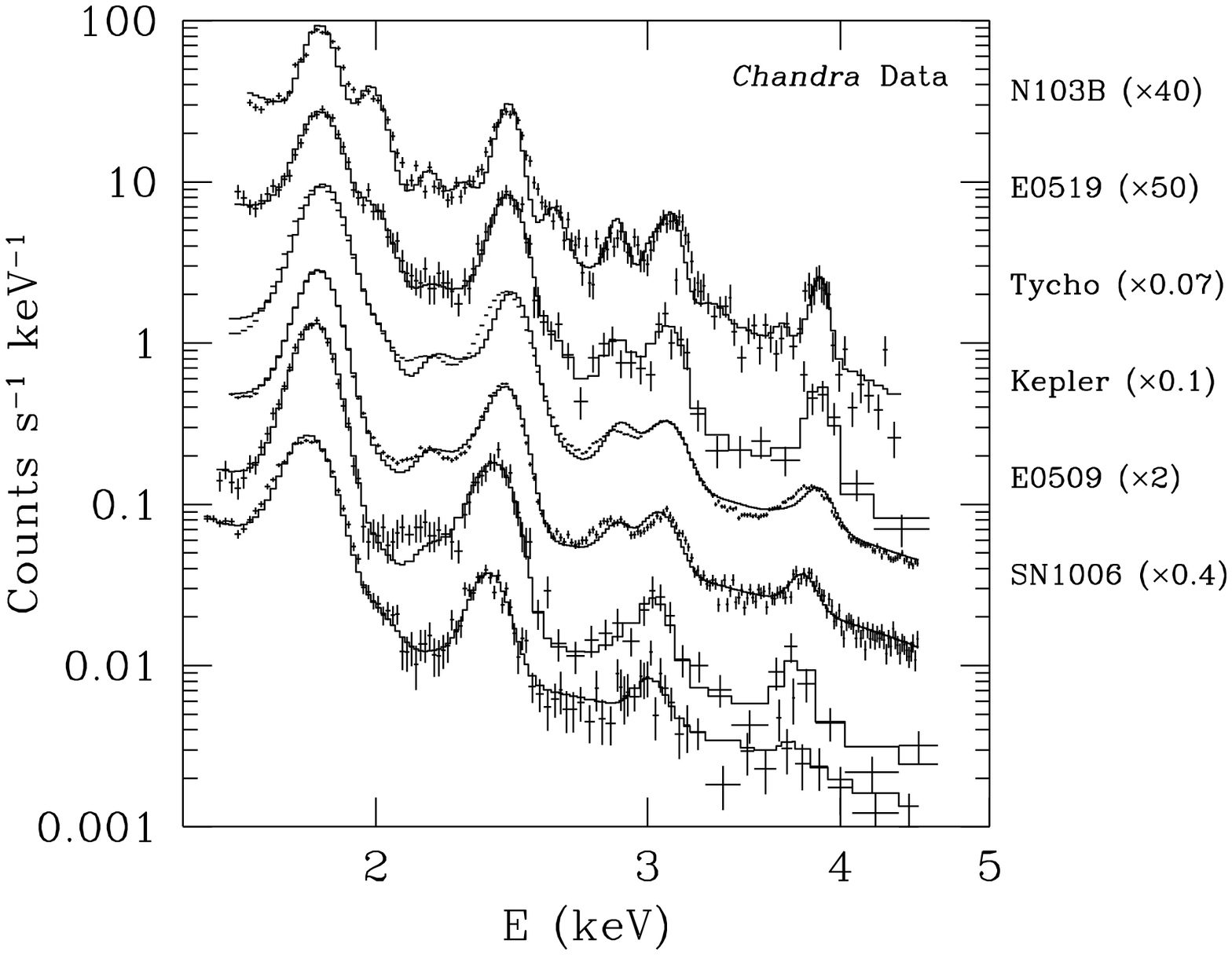}
  \includegraphics[scale=0.4]{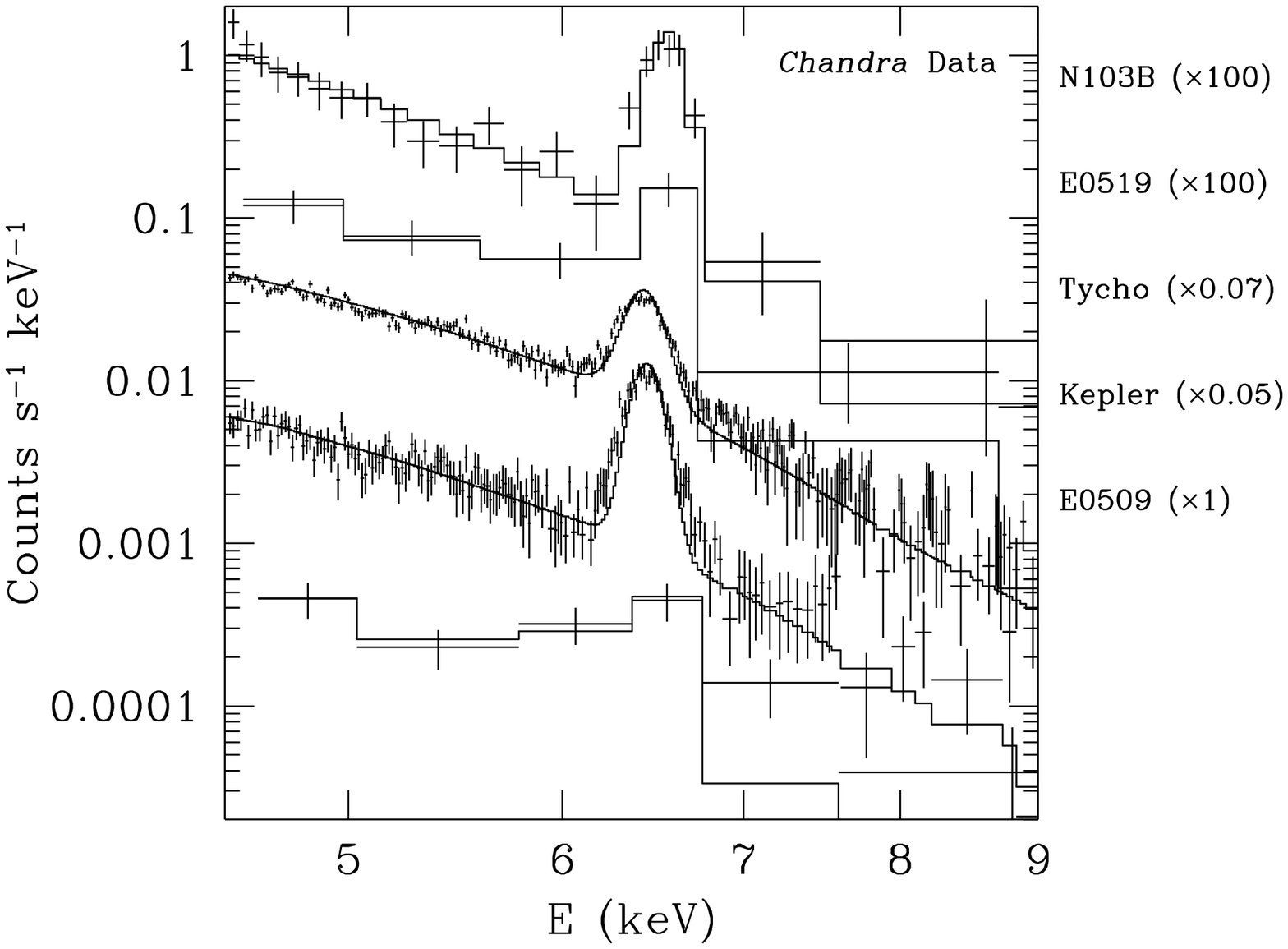}

  \caption{Plane-parallel shock fits to the \textit{Chandra} spectra of the SNRs in the sample. The K-shell emission
    blends from Si, S, Ar, and Ca are on the left panel, and the Fe K blend is on the right panel. The spectra are
    roughly ordered from highest (top) to lowest (bottom) ionization timescale. \label{fig-10}}

\end{figure}

In order to minimize these sources of error and determine values of $n_{e}t$ in different SNRs that can be compared to
each other (and to the models) with some confidence, we have downloaded archival \textit{Chandra} observations for the
objects in our sample, and we have fitted them with the same procedure. We used the plane-parallel shock models
described in \citet{hughes00:E0102} for the thermal component and a power law to describe the nonthermal continuum, both
of which were absorbed by an intervening column of Galactic interstellar matter assumed to be of solar
composition. During the fits, the Galactic column density was fixed to a value determined from the entire
\textit{Chandra} spectrum. To minimize the errors related to atomic data, we concentrated on the K-shell lines from the
most abundant elements (spectra above 1.5 keV). This has the advantage of reducing the impact of uncertainties in the
absorption as well. In fits to the 1.5--5 keV band (`Si component') we included emission from Si, S, Ar, and Ca, all
with the same $T_{e}$ and $n_{e}t$. The relative abundances of these species were allowed to vary during the fit, as
were the index and normaliztaion of the power-law. Independent fits were done to the 5--10 keV band around the Fe K
blend (`Fe component'). Only Fe and the power-law continuum (plus fixed absorption) were included in the Fe fits. The
best-fit models for each SNR in both bands can be seen in Figure \ref{fig-10}.
 
The relationship between the two fundamental quantities $n_{e}t$ and $T_{e}$ in these fits is of particular
interest. Plane-parallel shock models implicitly rely on the thermal continuum to constrain $T_{e}$ and the line
emission to constrain $n_{e}t$. Unfortunately, the X-ray continuum is often dominated by nonthermal emission (the
power-law in our fits), and the value of $T_{e}$ is largely unconstrained. We have taken this situation into account by
producing fits at fixed values of $T_{e}$: 1 and 10 keV for the Si component, 5 and 10 keV for the Fe component. These
values represent reasonable maxima and minima for shock-heated plasma in young SNRs, and the best-fit value for $T_{e}$
(models plotted in Figure \ref{fig-10}) is bounded by them in all cases. The range of ionization timescales listed in
Table \ref{tab-4} corresponds to these extrema. We have also subtracted 0.3 dex from the fitted values to account for
the fact that the plane-parallel shock models contemplate a distribution of ionization timescales between 0 and
$n_{e}t$, so the average ($\langle n_{e}t \rangle$) is one half of the fitted value.

A rough idea of the level of agreement between our hydrodynamic models and the spectral fits to the observations can be
obtained by comparing the fitted values of $log(\langle n_{e}t \rangle)$ in the Tycho SNR from Table \ref{tab-4}
(10.22-10.99 for Si, 9.72-9.78 for Fe) with the values of $log(\langle \tau \rangle)$ in the best model for the ejecta
emission found by \citet{badenes05:tycho} (10.12 for Si, 9.69 for Fe). These small deviations support the validity of
the comparisons made in \S~\ref{sub:x-ray-emission:modvsobs}.

We conclude with an unrelated (but important) remark on the fitted ionization timescales listed in Table \ref{tab-4}. In
all the Type Ia SNRs we have examined, the $\langle n_{e}t \rangle$ of Si is significantly higher than that of Fe, sometimes
by more than an order of magnitude. This constitutes strong evidence that the ejecta of Type Ia SNe must be stratified
to some degree (with most of the Fe interior to most of the Si), favoring delayed detonation models over deflagration
models with well-mixed ejecta \citep[see discussion in \S~3 of][]{badenes05:xray}.

\begin{deluxetable}{ccccccc}
  \tablewidth{0pt} 
  \tabletypesize{\scriptsize}
  \tablecaption{Outflow Models for Type Ia Progenitor Systems \label{tab-1}} 
  \tablecolumns{7} 
  \tablehead{ 
    \colhead{Model} &
    \colhead{$M_{of}$} & 
    \colhead{$t_{SN}$} &
    \multicolumn{3}{c}{Binary System Parameters} &
    \colhead{Reference}\\
    \colhead{Name} & 
    \colhead{$(\mathrm{M_{\odot}})$} & 
    \colhead{$(\mathrm{yr})$} &
    \colhead{$M_{WD,0}\,\mathrm{(M_{\odot})}$} &
    \colhead{$M_{D,0}\,\mathrm{(M_{\odot})}$} &
    \colhead{$P_{0}\,\mathrm{(days)}$} &
    \colhead{} 
  } 

  \startdata
  H1 & 0.15 & $5.0\times10^{5}$ & 1.0 & 2.0 & 2.0 & 1 (Fig. 7)\\
  LV1 & 0.50 & $1.8\times10^{6}$ & 1.0 & 2.5 & 1.6 & 2 (Fig. 1)\\
  HP1 & 0.24 & $2.0\times10^{6}$ & 0.75 & 2.0 & 1.58 & 3 (Fig. 1a)\\
  HP2 & 0.80 & $1.4\times10^{6}$ & 0.8 & 2.2 & 2.50 & 3 (Fig. 1c)\\
  HP3 & 0.50 & $6.0\times10^{5}$ & 1.0 & 2.4 & 3.98 & 3 (Fig. 1e)\\
  L1 & 0.40 & $2.0\times10^{6}$ & 1.0 & 2.3 & 1.74 & 4 (Model 2, Fig.7)\\
  L2 & 0.64 & $2.0\times10^{6}$ & 0.8 & 2.1 & 1.53 & 4,5 (Model 31, Fig. 36 in ref. 5)
  \enddata
  
  \tablecomments{Outflow model parameters: $M_{of}$: total mass lost by the system; $t_{SN}$: time between the onset of
    the final mass transfer episode and the SN explosion. Binary system parameters: $M_{WD,0}$: initial mass of the WD;
    $M_{D,0}$: initial mass of the donor; $P_{0}$: initial period.}

  \tablerefs{(1): \cite{hachisu99:supersoft}; (2): \cite{li97:supersoft}; (3):
    \cite{han04:SDchannel_for_SNIa}; (4): \cite{langer00:prog-models}; (5): \cite{deutschmann98:Iaprogenitors}}
  
\end{deluxetable}

\begin{deluxetable}{cccccccccc}
  \tablewidth{0pt} 
  \tabletypesize{\scriptsize}
  \tablecaption{Structure of the cavities produced by Type Ia Progenitor Outflows \label{tab-2}} 
  \tablecolumns{10} 
  \tablehead{
    \colhead{} &
    \multicolumn{4}{c}{Outflow Parameters} &
    \multicolumn{5}{c}{Cavity Parameters}\\
    \colhead{} &
    \colhead{$u_{of}$} &
    \colhead{$\rho_{ISM}$} &
    \colhead{$T_{ISM}$} &
    \colhead{Outflow} &
    \colhead{$R_{uof}$} &
    \colhead{$M_{uof}$} &
    \colhead{$R_{c}$}& 
    \colhead{$M_{sh}$} &
    \colhead{Stage} \\
    \colhead{Model} & 
    \colhead{$\mathrm{(cm \, s^{-1})}$} &
    \colhead{$\mathrm{(g \, cm^{-3})}$} &
    \colhead{$\mathrm{(K)}$} &
    \colhead{Regime} &
    \colhead{$\mathrm{(cm)}$} & 
    \colhead{$\mathrm{(M_{\odot})}$} & 
    \colhead{$\mathrm{(cm)}$} & 
    \colhead{$\mathrm{(M_{\odot})}$} &
    \colhead{at $t_{SN}$} 
  } 

  \startdata
  H1 & $10^{8}$ & $10^{-24}$ & $10^{4}$ & F & - & - & $5.25\times10^{19}$ & $3.05\times10^{2}$ & APCB\\
  LV1 & $10^{8}$ & $10^{-24}$ & $10^{4}$ & F & - & - & $9.75\times10^{19}$ & $1.95\times10^{3}$ & APCB\\
  HP1 & $10^{8}$ & $10^{-24}$ & $10^{4}$ & F & - & - & $8.30\times10^{19}$ & $1.20\times10^{3}$ & APCB\\
  HP2 & $10^{8}$ & $10^{-24}$ & $10^{4}$ & F & - & - & $1.06\times10^{20}$ & $2.51\times10^{3}$ & APCB\\
  HP3 & $10^{8}$ & $10^{-24}$ & $10^{4}$ & F & $1.40\times10^{19}$ & $8.00\times10^{-3}$ & $6.68\times10^{19}$ & $7.68\times10^{2}$ & ABROS\\
  L1 & $10^{8}$ & $10^{-24}$ & $10^{4}$ & F & $1.17\times10^{19}$ & $1.45\times10^{-3}$ & $9.23\times10^{19}$ & $1.66\times10^{3}$ & APCB\\
  L2 & $10^{8}$ & $10^{-24}$ & $10^{4}$ & F & - & - & $1.10\times10^{20}$ & $2.80\times10^{3}$ & APCB \\
  HP3lowp & $10^{8}$ & $5\times10^{-25}$ & $5\times10^{3}$ & F & $1.80\times10^{19}$ & $1.03\times10^{-3}$  & $7.82\times10^{19}$ & $5.97\times10^{2}$ & ABROS\\ 
  L2lowp & $10^{8}$ & $5\times10^{-25}$ & $5\times10^{3}$ & F & - & - & $1.35\times10^{20}$ & $2.59\times10^{3}$ & APCB\\
  HP3u1e6 & $10^{6}$ & $10^{-24}$ & $10^{4}$ & S & $3.75\times10^{18}$ & 0.28 & $4.89\times10^{18}$ & 1.64 & RB\\ 
  HP3u1e7 & $10^{7}$ & $10^{-24}$ & $10^{4}$ & S & $1.18\times10^{19}$ & $8.10\times10^{-2}$ & $1.70\times10^{19}$ & 0.65 & PRB\\ 
  HP3u2e7 & $2 \times 10^{7}$ & $10^{-24}$ & $10^{4}$ & F & $1.14\times10^{19}$ & $3.75\times10^{-2}$ & $3.13\times10^{19}$ & $1.44\times10^{2}$ & ABROS\\ 
  L2u1e6 & $10^{6}$ & $10^{-24}$ & $10^{4}$ & S & - & - & $9.80\times10^{17}$ \tablenotemark{a} & 2.45 \tablenotemark{a} & RB (C)\\
  L2u1e7 & $10^{7}$ & $10^{-24}$ & $10^{4}$ & S & - & - & - & - & PRB (C)\\ 
  L2u2e7 & $2 \times 10^{7}$ & $10^{-24}$ & $10^{4}$ & F & - & - & $3.13\times10^{19}$ & 52.5 & APCB
  \enddata

  \tablecomments{Outflow regime: fast (F) or slow (S). Cavity parameters: $R_{uof}$: radius of the unshocked outflow;
    $M_{uof}$: mass of the unshocked outflow; $R_{c}$: radius of the CD (for models in the fast outflow regime, this is
    equal to the \textit{inner} radius of the low-density cavity); $M_{sh}$: mass in the radiatively cooled shell (for
    cavities that are not pressure confined, $M_{sh}$ is the mass contained between $R_{c}$ and the outer supersonic
    blast wave). For the cavity stage at $t_{SN}$, we use the terminology of \citet{koo92:bubbles_II}. APCB: adiabatic
    pressure-confined bubble; ABROS: adiabatic bubble with a radiative outer shock; RB: radiative bubble; PRB: partially
    radiative bubble. Cavities marked with '(C)' have collapsed at $t_{SN}$.}
  
  \tablenotetext{a}{Parameters for the relic shell of the collapsed bubble.}

\end{deluxetable}

\begin{deluxetable}{cccccc}
  \tablewidth{0pt} 
  \tabletypesize{\scriptsize}
  \tablecaption{Type Ia Supernova Remnants - Forward Shock Dynamics\label{tab-3}} 
  \tablecolumns{6} 
  \tablehead{
    \colhead{Remnant} &
    \colhead{Age} &
    \colhead{$D$} &
    \colhead{$\alpha_{FS}$} &
    \colhead{$R_{FS}$} &
    \colhead{$u_{FS}$}\\
    \colhead{Name} &
    \colhead{(yr)} &
    \colhead{(kpc)} &
    \colhead{(arcmin)} &
    \colhead{(cm)} &
    \colhead{$\mathrm{(cm\,s^{-1})}$} 
  } 

  \startdata
  SN1885 & 121 \tablenotemark{a} & $785\pm30$ (1) \tablenotemark{b} & $6.7\times10^{-3}$ (1) & $(4.7 \pm 0.2) \times 10^{18}$ & -\\
  Kepler & 402 \tablenotemark{a} & $4.8\pm1.4$ (2) & $1.7$ & $(7.3 \pm 2.1) \times 10^{18}$ & $2.0\times10^{8}-2.5\times10^{8}$ (3)\\ 
  Tycho & 434 \tablenotemark{a} & $1.5-3.1$ (4) & $4.0$ & $(8.3 \pm 2.9) \times 10^{18}$ & $1.9\times10^{8}-2.3\times10^{8}$ (5)\\ 
  SN1006 & 1000 \tablenotemark{a} & $1.4-2.8$ (4) \tablenotemark{c} & $15.0$ & $(2.8 \pm 0.9) \times 10^{19}$ & $2.8\times10^{8}-3.0\times10^{8}$ (3)\\ 
  0509-67.5 & $400\pm120$ (7) \tablenotemark{d} & $50\pm1$ (8) \tablenotemark{e} & 0.24 \tablenotemark{f} & $1.1 \times 10^{19}$ & $>2.0\times10^{8}$ (5) \\
  0519-69.0 & $600\pm200$ (7) \tablenotemark{d} & $50\pm1$ (8) \tablenotemark{e} & 0.26 \tablenotemark{f} & $1.2 \times 10^{19}$ & $1.0\times10^{8}-1.9\times10^{8}$ (5) \\
  N103B & $860\pm400$ (7) \tablenotemark{d,g} & $50\pm1$ (8) \tablenotemark{e} & 0.23 \tablenotemark{f} & $1.0 \times 10^{19}$ & - 
  \enddata

  \tablecomments{SNR parameters: $D$: distance to the SNR; $\alpha_{FS}$: angular radius of the FS; $R_{FS}$: FS radius
    ($R_{FS}=3.09 \times 10^{21} \, D (\alpha_{FS} / 60) (\pi / 180) $); $u_{FS}$: FS velocity.}

  \tablenotetext{a}{Historical SNR.} 

  \tablenotetext{b}{Distance to M31.}

  \tablenotetext{c}{For the distance to SN1006, we use the more conservative estimate of
    \citet{smith91:six_balmer_snrs}, which includes systematic as well as statistical uncertainties, instead of the
    newer results of \citet{winkler03:SN1006_Distance}.}

  \tablenotetext{d}{Age estimate from light echoes.}

  \tablenotetext{e}{Distance to the LMC.}

  \tablenotetext{f}{For the LMC SNRs, the radii have been determined by the outermost extent of the X-ray emission in
    the 0.5-4 keV band, subtracting 0.5'' due to the smearing of the rim by the \textit{Chandra} PSF.}

  \tablenotetext{g}{The uncertainty in the age of N103B arbitrarily set to 400 yr. \citet{rest05:LMC_light_echoes}
    determined an age of 860 yr, but could not constrain the uncertainty in the measurement.}
  
  \tablerefs{(1) \citet{fesen06:SN1885}; (2) \citet{reynoso99:distance_Kepler}; (3) \citet{sollerman03:balmer_shocks};
    (4) \citet{smith91:six_balmer_snrs}; (5) \citet{ghavamian01:balmer_SNRs}; (7) \citet{rest05:LMC_light_echoes}; (8)
    \citet{alves04:LMC_Distance}}

\end{deluxetable}

\begin{deluxetable}{ccccc}
  \tablewidth{0pt} 
  \tabletypesize{\scriptsize}
  \tablecaption{Type Ia Supernova Remnants - Ionization Timescales in the Shocked Ejecta \label{tab-4}} 
  \tablecolumns{4} 
  \tablehead{
    \colhead{Remnant} &
    \multicolumn{2}{c}{log $\langle n_{e}t \rangle$ ($\mathrm{cm^{-3}\,s}$)}\\
    \colhead{Name} &
    \colhead{Si} &
    \colhead{Fe} 
  } 

  \startdata
  Kepler & $10.08 - 10.24$ & $9.85 - 9.92$\\
  Tycho & $10.23 - 10.99$ & $9.72 - 9.78$\\
  SN1006 & $9.49 - 9.60$ & - \tablenotemark{a}\\
  0509-67.5 & $9.80 - 9.82$ & $<9.7$\\
  0519-69.0 & $10.50 - 11.62$ & $9.90 - 9.95$\\
  N103B & $10.64 - 11.94$ & $10.62 - 10.69$ 
  \enddata

  \tablenotetext{a}{SN 1006 has no Fe K emission in its \textit{Chandra} X-ray spectrum (Hughes et al., in preparation).}
  
\end{deluxetable}

\end{document}